\documentclass[aps,prb,amssymb,showpacs,twocolumn,floatfix]{revtex4-2}

\usepackage{graphics,graphicx,amsmath}
\usepackage{tabularx,float}
\usepackage{epsfig}
\usepackage{subfigure}
\usepackage{bm}
\usepackage{wasysym}
\usepackage{bbm}
\usepackage[T1]{fontenc}
\usepackage[normalem]{ulem}

\usepackage{mathrsfs} 
\usepackage{physics}

\usepackage{color}
\usepackage{verbatim}
\graphicspath{{fig/}}

\begin{document}

\date{\today}

\title{Normal state quantum geometry and superconducting domes in (111) oxide interfaces}

\author{Florian Simon\email[Corresponding author: ]{florian.simon1@universite-paris-saclay.fr}}
\affiliation{Laboratoire de Physique des Solides, Universit\'e Paris-Saclay,
CNRS UMR 8502, F-91405 Orsay Cedex, France} 
\author{Mark O. Goerbig}
\affiliation{Laboratoire de Physique des Solides, Universit\'e Paris-Saclay,
CNRS UMR 8502, F-91405 Orsay Cedex, France}
\author{Marc Gabay}
\affiliation{Laboratoire de Physique des Solides, Universit\'e Paris-Saclay,
CNRS UMR 8502, F-91405 Orsay Cedex, France}

\begin{abstract}

We theoretically investigate the influence of the normal state quantum geometry on the superconducting phase in (111) oriented oxide interfaces and discuss some of the implications for the $\text{LaAlO}_3/\text{SrTiO}_3$ (LAO/STO) heterostructure. From a tight-binding modeling of the interface, we derive a two-band low-energy model, allowing us to analytically compute the quantum geometry and  giving us access to  the superfluid weight, as well as to showcase the role of two particular relevant energy scales. One is given by the trigonal crystal field which stems from the local trigonal symmetry at the interface, and the other one is due to orbital mixing at the interface. Our calculations indicate that the variation of the superfluid weight with the chemical potential $\mu$ is controlled by the quantum geometry in the low-$\mu$ limit where it presents a dome. At higher values of $\mu$ the conventional contribution dominates. In order to  make quantitative comparisons between our results and experimental findings, we rely on an experimentally observed global reduction of the superfluid weight that we apply to both the conventional and geometric contributions. Furthermore, an experimentally measured non-monotonic variation of $\mu$ with the gate voltage $V_g$ is taken into account and yields a two-dome scenario for the superconducting critical temperature as a function of $V_g$. The observed dome in the low-$V_g$ regime is explained by the non-monotonic evolution of a dominant conventional part of the superfluid density. In contrast, the expected second dome at larger values of $V_g$ limit would be due to a dominant quantum-geometric contribution.

\end{abstract}
\maketitle

\section{Introduction}

Superconductivity has, since 1911, become a flagship of condensed-matter physics. The main paradigm is given by the Bardeen-Cooper-Schrieffer (BCS) theory~\cite{bardeen_theory_1957} which, in its standard form, describes the mechanism whereby quasiparticles in a single, partially filled band, pair and
condense in a single collective dissipationless state. This single-band approximation has its limits. Indeed, since the 1950s \cite{Adams1959,Blount1962},
it has been realized that in a multiband situation, even in the adiabatic limit, each band carries a signature of the other bands
in the form of two geometric contributions, namely in what has later been identified as the Berry curvature and the quantum metric \cite{berry_quantum_1989}. These quantities form what we call band/quantum geometry. In the context of superconductivity, this means that even if the Cooper pairing takes place within a single band, it is \textit{a priori} affected by the other electronic bands of the normal state, particularly through the \textit{normal state quantum geometry}. While BCS theory does not take these geometric effects into account, recent studies have theoretically pointed out the relevance of the quantum metric for the superfluid weight of flat-band models \cite{Peotta2015,liang_band_2017,Iskin2018,rossi_quantum_2021,torma_superconductivity_2022}, as well as of the Berry curvature of Dirac-like systems \cite{simon_role_2022}, such as 2D transition metal dichalcogenides. We note that the geometry of singlet superconductivity was studied in Ref.\cite{Porlles2023}.

Our study emphasizes the impact of the normal state quantum geometry on superconductivity for (111)-oriented oxide interfaces, and more specifically for the LaAlO$_3$-SrTiO$_3$ (LAO/STO) heterostructure \cite{Gariglio2016}. Let us point out that the results which we present here may be relevant for other materials, including other (111) oxide interfaces. The  LAO/STO heterostructure hosts a two-dimensional electron gas (2DEG) on the STO side, confined to a few layers in the vicinity of the interface \cite{Song2018}. For the (111) interface, carriers in the 2DEG move on a honeycomb structure with three orbitals per site and, from that point of view, this may be seen as a three-orbital version of graphene \cite{Doennig2013}. 

Starting from a tight-binding (TB) modeling of the interface, we develop a low-energy model around the $\Gamma$ point in two steps. First, we consider the spinless case for which the system has six bands. The dominance in energy of intra-orbital nearest-neighbor hoppings allows us to decouple the six bands into a bonding triplet and an anti-bonding triplet of bands. Focusing on the bonding triplet, we derive a low-energy three band model to quadratic order in $\mathbf{k}$. The latter presents two upper bands, degenerate at the $\Gamma$ point, that represent the $|e_{\pm g}\rangle$ states. The other band is lower in energy, because of the trigonal crystal field, and represents the $|a_{1g}\rangle$ state. The three bands have isotropic and quadratic dispersions but the lowest band is significantly heavier than the other two. The latter presents a strong peak of the quantum metric, driven by inter-orbital nearest neighbor hoppings, which we refer to as \textit{orbital mixing}. Second, we consider the spinful case in which a spin-orbit coupling term is taken into account. We then have a twelve-band TB model, from which we also derive a two-band low-energy model. The latter reveals an isotropic Dirac cone at the $\Gamma$ point, which stems from the orbital mixing. 

From the low-energy model, we can then draw a qualitative scenario for the chemical potential ($\mu$) dependence of the conventional and geometric contributions to the superfluid weight \cite{liang_band_2017,torma_superconductivity_2022}. The conventional contribution is essentially linear in $\mu$, because of the isotropy of the dispersion, and the geometric contribution presents a dome as a function of $\mu$, since the Dirac cone produces a strong peak in the quantum metric. 

Finally, we compare the results of the qualitative scenario to the experimental data, in the case of the LAO/STO interface. Taking into account thermal and disorder effects, we compute the superfluid weight  both for the low-energy model, and the spinful TB model. In both cases, we find the correct order of magnitude for the superfluid weight (and the associated BKT temperature). The most significant difference between the two models is the value of $\mu$ below which the geometric contribution becomes dominant. In the low-energy model, the $\mu$ functional form of this quantity displays such  a steep variation that it seems hardly possible to experimentally observe  the regime when it dominates. In the spinful TB model, bands from the anti-bonding group, though $6-7$eVs away, contribute significantly to the quantum metric of the lowest Kramers' partners, producing a smoother variation of the geometric term and increasing the value of chemical potential below which the geometrical contribution dominates. 

While the evolution of the superfluid weight is naturally discussed in terms of the chemical potential within a theoretical approach, experiments use the gate voltage $V_g$ as a tuning parameter rather than $\mu$. Oftentimes the variation of $\mu$ with $V_g$ is monotonic both quantities are related by a monotonic function but Hall transport experiments \cite{Monteiro2019,Khanna2019} have indicated that this is not the case for the LAO/STO (111) interfaces, where a non-monotonic dependence of the Hall carrier density on $V_g$ has been observed. Assuming that the Hall carrier density is a monotonic function of the chemical potential, this suggests that the chemical potential is non monotonic in $V_g$. Taking this as an experimental fact, we then infer a qualitative dependence of the superfluid weight (and thus BKT temperature) on the gate voltage. The experimentally observed dome is explained by  the conventional contribution and the non monotonic dependence of $\mu$ on $V_g$. As $V_g$ is further increased, we expect a saturation effect followed by
the appearance of a \textit{second superconducting dome}, due purely to the geometric contribution to the superfluid weight. 

The paper is organized as follows. In section \ref{sec:TB}, we present the tight-binding model, both in the spinless and spinful cases. In section \ref{sec:low_energy_model}, we derive low-energy models and discuss their structure, both in the spinless and the spinful cases. We then build on the spinful low-energy model to discuss the dependence of the superfluid weight on the chemical potential. Finally, in section \ref{sec:semiquantitative SFW} we make contact with experiments for the spinful case. We numerically compute the superfluid weight versus the chemical potential, for the low-energy model and then for the full TB model, and we discuss the extent to which the results that are obtained are consistent with the  scenario presented in section \ref{sec:qualitative SFW}.

\section{Tight-binding model}\label{sec:TB}

\begin{figure}[ht!]
    \centering
    \includegraphics[width=\columnwidth]{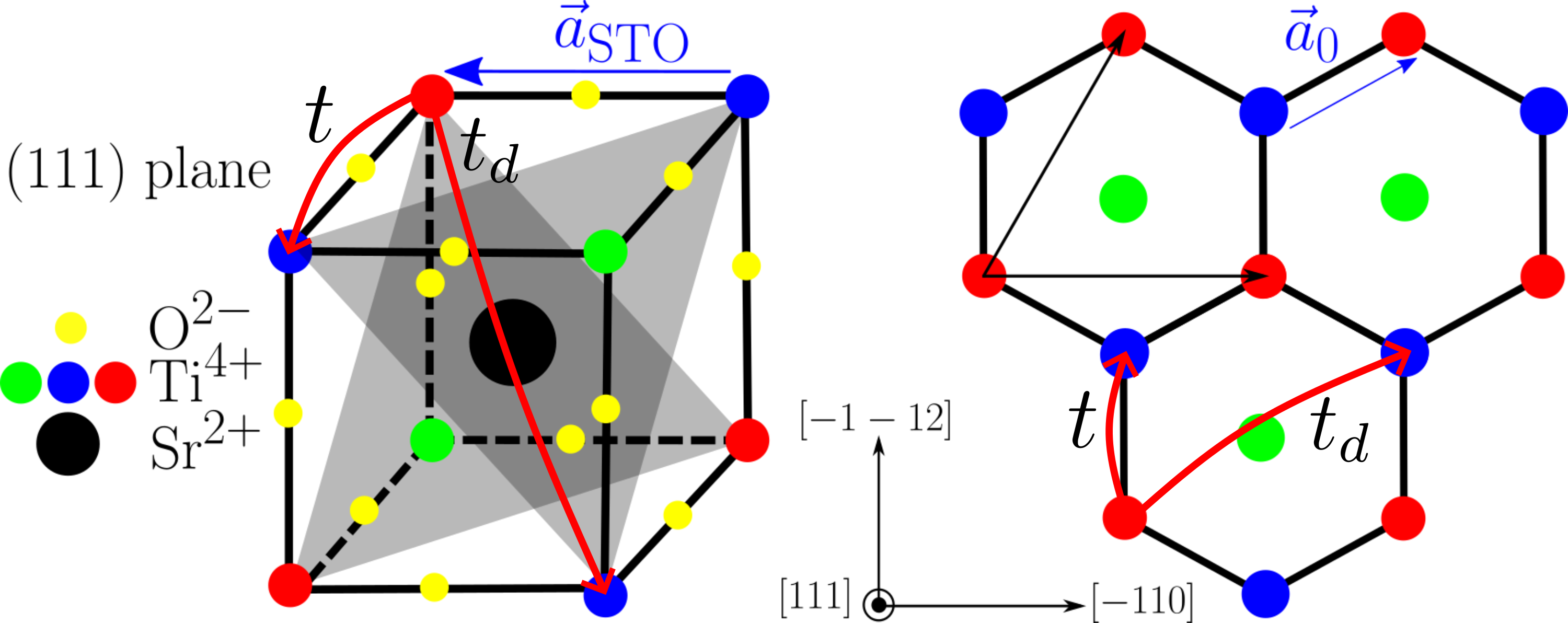}
    \caption{STO side, just below the (111) LAO/STO interface, adapted from \cite{rodel_orientational_2014}. Left: cubic lattice cell the corners of which are occupied by Ti$^{4+}$ ions. The gray areas indicate  planes  normal to the [111] direction. Right: projection onto (111) planes. Two layers of Ti$^{4+}$ ions (blue and red) form a honeycomb lattice, where the two triangular sublattices are displaced by the vector $\mathbf{a}_0$. Lastly, $a_{\text{STO}}=3.905\text{\AA}$ and $a_0=\sqrt{2/3}a_{\text{STO}}$. Two different hoppings are pictured. A nearest-neighbor hopping with strength $t$, and a third nearest-neighbor hopping with strength $t_d$. They lead to a band structure in quantitative agreement with ARPES spectra}
    \label{Schema_LAOSTO}
\end{figure}

We first introduce the relevant TB modeling of the $(111)$ interface and discuss its various terms. The values of the relevant energy scales, presented in detail in Sec. \ref{sec:low_energy_model}, are mainly taken from Refs. \cite{rodel_orientational_2014,de_luca_symmetry_2018,Vivek2017,Song2018,Khanna2019}. The structure of the (111) oriented STO substrate is shown in Fig. \ref{Schema_LAOSTO}.
 The two-dimensional electron gas (2DEG) is located on the STO side of the LAO/STO interface \cite{Song2018}. From a structural point of view, the three-dimensional (3D) SrTiO$_3$ crystal has a ABO$_3$ cubic perovskite structure (left panel, Fig. \ref{Schema_LAOSTO}). In the (111) orientation, see Fig. \ref{Schema_LAOSTO},  two consecutive
(111) planes contain Ti ions for one and SrO$_3$ ionic groups
for the other. Focusing on the Ti (111) planes, the atomic arrangement consists of layers of two-dimensional (2D) triangular lattices displaced by the vector $\mathbf{a}_0$ (see Fig. \ref{Schema_LAOSTO}). Consequently, the Ti atoms form ABC-stacked two-dimensional 2D triangular lattices in the (111) planes (see Fig. \ref{Schema_LAOSTO}, left panel). From an electronic point of view, the charge carriers hop precisely between neighboring Ti atoms through direct orbital overlap or via the O sites. 

While the basic building block for the description of the 2DEG would in principle contain three (ABC) layers of Ti atoms (red, blue and green in Fig. \ref{Schema_LAOSTO}), the location of the Fermi energy for the (111) direction, as seen in Ref. \cite{Song2018}, allows us to reduce the model to only two layers shown in Fig. \ref{Schema_LAOSTO} (right panel), as we have checked explicitly numerically in Appendix \ref{Annexe_bilayer}. Indeed, for a unit trilayer stack, the TB Hamiltonian describing the kinetics of the 2DEG parallel to the (111) interface produces nine bands (not counting spin) organized in three  groups (bonding, non-bonding, anti-bonding). Numerical inspection, using representative values of hopping amplitudes, shows that the energy difference between consecutive groups is on the order of several eV such that the three non-bonding bands which come from the third (green) layer are several eVs away from the Fermi energy, as discussed in Appendix \ref{Annexe_bilayer}. Additionally, the dispersions of the occupied bonding triplet bands show very little difference with those of a bilayer model, where only the first (blue) and second (red) layers are considered  (see Fig. \ref{Compare23couches} of Appendix \ref{Annexe_bilayer}). 

We can therefore leave out the third layer and consider the system shown on the right in Fig. \ref{Schema_LAOSTO}, \textit{i.e.} two triangular layers (red and blue) displaced by the vector $\mathbf{a}_0$ that form a honeycomb lattice characterized by the layer/sublattice index $\big\{1,2\big\}$.
On each site, we have the three conducting $t_{2g}$ Ti orbitals, $\big\{d_{yz},d_{xz},d_{xy}\big\}$, which, in the following, we denote $\big\{\text{X},\text{Y},\text{Z}\big\}$, respectively. 

Accounting for spin, we then have a honeycomb lattice with two spins and three orbitals per site,  yielding a twelve-band system. Next, we discuss each type of hopping entering our TB model.
    \subsection{Kinetic term}    
The kinetic part of the model takes into account hoppings between the different lattice sites and orbitals. This term  describes carrier motions conserving the orbital character, with amplitudes $t$ and $t_d$ for nearest and third nearest neighbors belonging to two different layers, respectively. The general form of the kinetic term is thus diagonal in terms of the orbitals but off-diagonal in terms of the layers. Therefore, in the $\{1,2\}\otimes\big\{\text{X},\text{Y},\text{Z}\big\}$ basis the kinetic term reads 
    \begin{equation}
        \begin{pmatrix}
            0&H_{\text{cin}}\\
            H_{\text{cin}}^*&0
        \end{pmatrix}=\tau_x\otimes\text{Re}(H_{\text{cin}})-\tau_y\otimes\text{Im}(H_{\text{cin}}),
        \label{kinetic_6}
    \end{equation}
with $H_{\text{cin}}=t\operatorname{diag}(e,f,g)$ in the orbital subspace. The Pauli matrices $\tau_x$ and $\tau_y$ in Eq. (\ref{kinetic_6}) act on the layer subspace. Explicit expressions for the functions $e,f$ and $g$ may be found in appendix \ref{expressions cinetiques}.

    \subsection{Orbital mixing terms}
While the kinetic term does not couple different orbitals, such couplings are generated at the interface by \textit{orbital mixing}. In appendix \ref{annexe OM}, we show by symmetry considerations that a natural choice is 
    \begin{equation}
        \tau_x\otimes H_{\text{om}}=\tau_x\otimes c_0\begin{pmatrix}
           0&i\delta&-i\alpha\\
           -i\delta&0&i\beta\\
           i\alpha&-i\beta&0
        \end{pmatrix},
        \label{orbital mixing matrix}
    \end{equation} 
    where $\alpha=\sin(\sqrt{3}/2k_x+3/2k_y)$, $\beta=\sin(\sqrt{3}/2k_x-3/2k_y)$, $\delta=-\sin(\sqrt{3}k_x)$ and $c_0$ the strength of the orbital mixing.  Here, we measure the wave vectors in units of the inverse $a_0^{-1}$ of the distance between nearest-neighbor sites in the (111) plane (see Fig. \ref{Schema_LAOSTO}), and $\tau_x$ is again a Pauli matrix acting on the layer degree of freedom. Note that with inversion symmetry, these terms are prohibited. But in reality, interfaces between $\text{LaAlO}_3$ and $\text{SrTiO}_3$ always have corrugation \cite{khalsa_theory_2013,zhong_theory_2013}, such that inversion symmetry is broken and orbitals that would have been orthogonal are not, resulting in non-zero overlap and allowed interorbital hoppings. It will give rise to an orbital Rashba effect. 

    \subsection{Trigonal crystal field}
Note that the (111) interface has a different point group symmetry than the orbitals whose symmetry is governed by the (cubic) bulk symmetry of LAO and STO. Therefore the $t_{2g}$ orbitals are not orthogonal to each other in the hexagonal lattice, resulting in a \textit{trigonal crystal field}, where the couplings have the same value because of the hexagonal symmetry. It lifts the degeneracy between the $e_{\pm g}$ orbitals and the $a_{1g}$ orbital within the conducting $t_{2g}$ orbitals of Ti. This trigonal crystal field, of strength $d$, thus couples the different orbitals in the same layers so that it may be written as 
\begin{equation}
    H_d=-d\tau_0\otimes \begin{pmatrix}
        0&1&1\\
        1&0&1\\
        1&1&0
    \end{pmatrix},
    \label{crystal field matrix}
\end{equation}
where $\tau_0$ is the identity matrix indicating that the trigonal crystal field is diagonal in the layer index. 

    \subsection{Confinement energy}
Finally, we need to take into account a confinement term that reflects the different onsite potentials for the two sublattices, which reside in different layers. 
It is equivalent to the Semenoff mass in graphene, breaking the $\mathcal{C}_6$ symmetry down to $\mathcal{C}_3$. We have $-V\Lambda_0$ for layer 1 and $V\Lambda_0$ for layer 2, so that this term may be written as $\tau_z\otimes(-V\Lambda_0)$, in terms of the $3\times 3$ identity matrix noted $\Lambda_0$. While this term may be important for other properties of the LAO/STO interface, we will see that it does not affect those studied in this paper, and we will later omit it when reducing the TB model to a low-energy model.

    \subsection{Spinless six-band model}
With these four terms, the six-band TB model is written, in $\{1,2\}\otimes\big\{\text{X},\text{Y},\text{Z}\big\}$, as
\begin{equation}
    H_6=\begin{pmatrix}
        -VI_3+H_d&H_{\text{cin}}+H_{\text{om}}\\
        H_{\text{cin}}^*+H_{\text{om}}&VI_3+H_d
   \end{pmatrix}.
    \label{six band matrix}
\end{equation}
A more convenient basis is the \textit{trigonal basis} in which the trigonal crystal field term is diagonal. The latter is detailed in appendix \ref{trigonal basis}. Hereafter, we discuss the band structure described by $H_6$ in the trigonal basis. We now add spin to our problem.
        \subsection{Spin-orbit coupling}
        The spin-orbit coupling (SOC) term is identical for the two layers,  and thus naturally diagonal in the layer subspace. In the $\{\uparrow,\downarrow\}\otimes\{\text{X},\text{Y},\text{Z}\}$ basis, it reads \cite{Khalsa2012}
        \begin{equation}
            H_{\text{SOC}}=-\lambda\sigma_x\otimes\Lambda_7+\lambda\sigma_y\otimes\Lambda_5-\lambda\sigma_z\otimes\Lambda_2,
        \end{equation}
        where the $\Lambda$-matrices denote the Gell-Mann matrices, explicited in appendix \ref{annexe:Gell-Mann matrices}. The energy scale $\lambda$ is given by $\lambda\simeq8$meV \cite{Khanna2019}.
        \subsection{Spinful twelve-band model}
         The spinless Hamiltonian $H_6$ is identical for each spin, the corresponding term in the $\{\uparrow,\downarrow\}\otimes\{1,2\}\otimes\big\{\text{X},\text{Y},\text{Z}\big\}$ basis, will be $\sigma_0\otimes H_6$. We thus finally obtain our twelve-band TB model, as
         \begin{align}
             H_{12}&=\sigma_0\otimes H_6-\lambda\sigma_x\otimes\tau_0\otimes\Lambda_7+\lambda\sigma_y\otimes\tau_0\otimes\Lambda_5\nonumber\\&-\lambda\sigma_z\otimes\tau_0\otimes\Lambda_2,
         \end{align}
 where the last three terms correspond to $\tau_0\otimes H_{\text{SOC}}$.
\section{Low-energy model}
\label{sec:low_energy_model}
We now aim to derive a low-energy expression from the TB model. We first discuss the problem without spin, and derive a low-energy three-band model, valid for each spin orientation. We then add the spin degree of freedom  and find that the low-energy limit leads to a two-band model. To do so, we only apply momentum independent unitary transformations, that leave the quantum geometry invariant.

    \subsection{Spinless problem}
Numerical diagonalization shows that the low-filling regime occurs near the $\Gamma$ point. Moreover, in the vicinity of the latter, there are two groups of three bands separated by several eV. This is because the gap between the two groups at the $\Gamma$ point is $2(2t+t_d)\sim6.5$eV, and the kinetic energy 
is clearly the largest energy scale. Therefore, for low fillings, it appears possible to reduce the above six-band expression to two effective three-band models, one for each group. To make a similar structure appear explicitly in $H_6$, we apply the following unitary transformation
\begin{equation}
    U=U_l\otimes\Lambda_0=\frac{1}{\sqrt{2}}\begin{pmatrix}
        -1&1\\
        1&1
    \end{pmatrix}\otimes\Lambda_0,
    \label{Transformation_spinless_case}
\end{equation}
which maximally entangles the two layers, in a symmetric $|s_l\rangle$ and an anti-symmetric combination $|a_l\rangle$. The six-band Hamiltonian is then transformed to
\begin{widetext}
    \begin{equation}
        U^\dagger H_6U=\begin{pmatrix}
        H_d-H_{\text{om}}-\text{Re}(H_{\text{cin}})&-VI_3+i\text{Im}(H_{\text{cin}})\\
        -VI_3-i\text{Im}(H_{\text{cin}})&H_d+H_{\text{om}}+\text{Re}(H_{\text{cin}})
    \end{pmatrix}.
    \label{H transforme}
    \end{equation}
\end{widetext}
Numerical comparison of the band dispersions of the diagonal blocks in Eq.(\ref{H transforme}) and that of the full spinless TB Hamiltonian in Eq.(\ref{six band matrix}) confirms that the former accurately reproduce the two  groups. Thus, we may focus on the lower diagonal block, which corresponds to the symmetric entanglement of the two layers, and take it as a low-energy three-band model that reads 
\begin{equation}
    H_3=H_d+H_{\text{om}}+\text{Re}(H_{\text{cin}}).
\end{equation}
A discussion of the validity of this approximation, done in appendix \ref{validity annexe}, shows that with a precision of a few meV, this \textit{three-band approximation} is valid over an area centered at $\Gamma$ and covering approximately ten percent of the Brillouin zone (BZ). To be consistent with this approximation, we need to expand $H_3$ to quadratic order in $k$.

        \subsubsection{Quadratic three-band model}
    
In appendix \ref{annexe DL quad H3}, we show that to quadratic order, we have 
    \begin{widetext}
    \begin{equation}
    H_3=-(2t+t_d)\Big(1-\frac{1}{4}k^2\Big)\Lambda_0+\begin{pmatrix}
        d-t_{\text{eff}}(k_x^2-k_y^2)&-2t_{\text{eff}}k_xk_y&ick_x\\
        -2t_{\text{eff}}k_xk_y&d+t_{\text{eff}}(k_x^2-k_y^2)&ick_y\\
        -ick_x&-ick_y&-2d
    \end{pmatrix},
    \label{H3quad}
\end{equation}
\end{widetext}
with $t_{\text{eff}}=(t-t_d)/8$ and $c=3c_0/\sqrt{2}$. Note that $H_3$ is expressed in the trigonal basis (see appendix \ref{trigonal basis}). The trigonal crystal field lifts the threefold degeneracy at the $\Gamma$ point (between $a_{1g}$ and $e_{\pm g}$ states). The linear and quadratic terms arise from the orbital mixing and kinetic terms, respectively. Inspection of Eq. (\ref{H3quad}), reveals that the $a_{1g}$ state is only coupled to the $e_{\pm g}$ states through the orbital mixing contribution, which generates its quantum geometry. 

$H_3$ can then be exactly diagonalized, and we find the following eigenvalues for the last term:
\begin{equation}
    \epsilon_1=d+t_{\text{eff}}k^2,\quad\epsilon_2=d+\bigg(\frac{c^2}{3d}-t_{\text{eff}}\bigg)k^2,
\end{equation}
and
\begin{equation}
    \epsilon_3=-2d-\frac{c^2}{3d}k^2,
\end{equation}
to quadratic order in the wave-vector components. The values taken hereafter are those corresponding to Refs. \cite{rodel_orientational_2014,de_luca_symmetry_2018,Vivek2017,Song2018,Khanna2019}, i.e. $t=1.6$ eV, $t_d=70$ meV, $V=100$ meV, $d=3$ meV. Additionally, we estimate $c_0=40$ meV.
We thus find an isotropic electron-like band structure. 

We point out that these results may apply to other (111) oxide interfaces. In Section \ref{sec:semiquantitative SFW}, we discuss the relevance of our results to the experimental context, illustrated with the case of the LAO/STO (111) interface. The lowest energy band ($\epsilon_3$) is substantially flatter than the other two. Indeed, its band mass can be computed to be 
\begin{equation}
    m_B=\frac{\hbar^2}{2a_0^2}\bigg(\frac{2t+t_d}{4}-\frac{c^2}{3d}\bigg)^{-1}\simeq21m_0,
    \label{band mass}
\end{equation}
with $m_0\simeq9.1\times10^{-31}$ kg the rest mass of an electron. This band presents a peak in its quantum metric at the value given by $g_{3,\mu\mu}(\Gamma)=c_0^2/9d\simeq90a_0^2$.

Note that going beyond the low-energy model, which is obtained using a quadratic order expansion in k,  we find that the interorbital contributions to cubic order, give rise to an orbital Rashba effect which moves the minimum away from the $\Gamma$ point and therefore the actual band mass differs from Eq. (\ref{band mass}).
We then plot this band structure and contrast it with the one we get from the TB form of the kinetic and orbital mixing terms in Fig. \ref{Structure_bandes}. The band structure of the full TB model in the full Brillouin Zone (BZ) is shown in Appendix \ref{annexe_band_struc_TB}. 
\begin{figure}[ht!]
    \centering
    \includegraphics[width=\columnwidth]{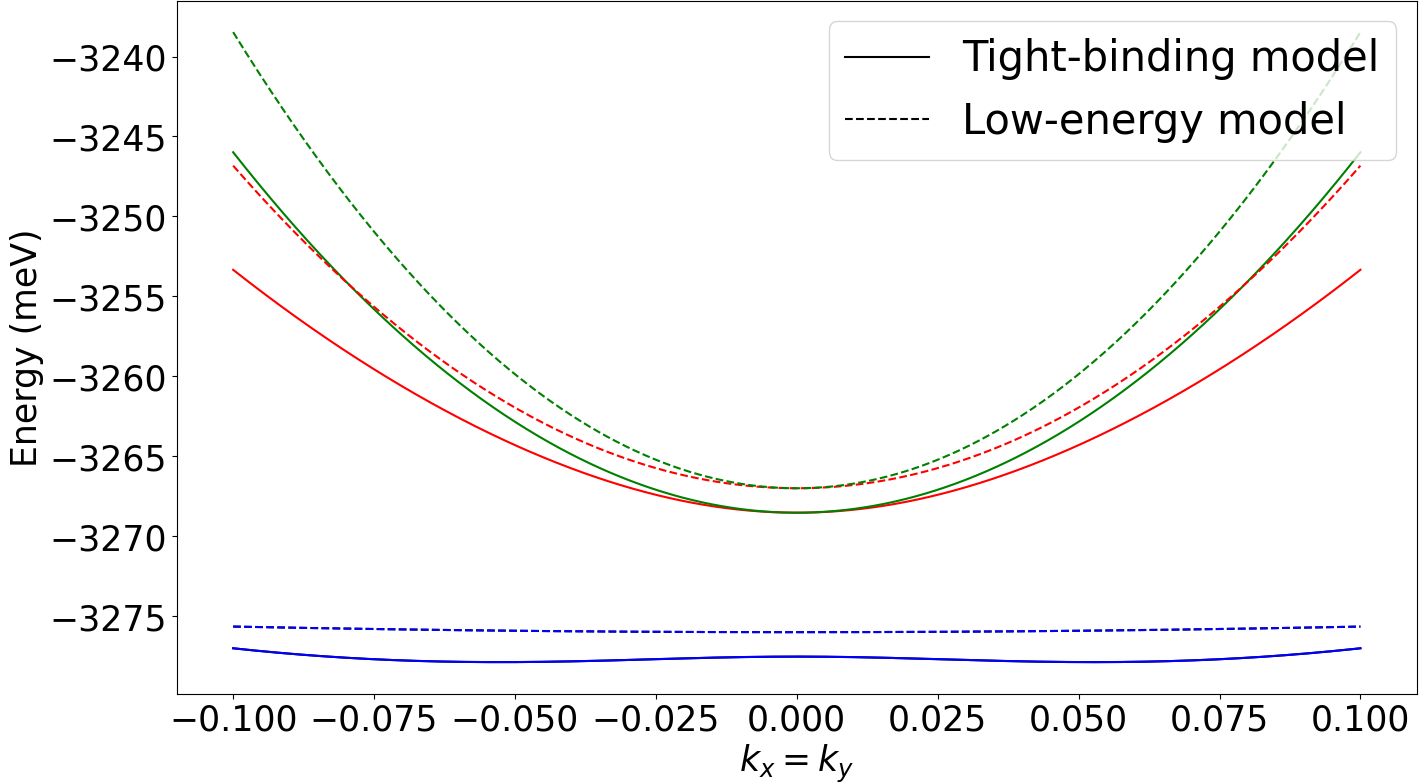}
    \caption{Energy dispersions near $\Gamma$, in the $k_x=k_y$ direction. Each color corresponds to one of the three bands, blue is the lowest band, red is the second lowest and green the third lowest band. Dashed lines correspond to the dispersions coming from the low-energy model, Eq. (\ref{H3quad}). Solid lines come from the full Tight-binding Hamiltonian, Eq. (\ref{six band matrix}).}
    \label{Structure_bandes}
\end{figure}    
We indeed get the aforementioned precision of a few meVs. Note that the offset of 2 meV between the TB and low-energy bands as seen in Fig. \ref{Structure_bandes}, is due to the fact that $V$ does not enter the low-energy expression of the Hamiltonian. Such a global shift does not have a physical relevance on the quantum geometry and superfluid weight as it can be compensated by a redefinition of the chemical potential with respect to the $\Gamma$ point value of the lowest band. We then get a lower band that is substantially flatter than the other ones and that is close in energy to a level crossing at the $\Gamma$ point.
    \subsection{Spinful problem}
We now derive a low-energy model starting from the spinful Hamiltonian $H_{12}$.
        \subsubsection{Derivation of the low-energy model}
We start by applying, once again, the unitary matrix $U$ given in Eq. (\ref{Transformation_spinless_case}) , and we use the matrix $P$, explicited in appendix \ref{trigonal basis}, to express the orbital part in the trigonal basis. The SOC term is left unchanged by $U$, while the $\sigma_0\otimes H_6$ term gets us two copies of Eq. (\ref{H transforme}). We then make the same approximation, and by switching the layer and spin degrees of freedom, we get a Hamiltonian which is the counterpart of the bonding and anti-bonding groups of the spinless case, only now with one copy for each spin as well as the accompanying SOC. We then restrict $H_{12}$ to the bonding subspace, i.e. the symmetric combination of the two layers $|s_l\rangle$, and then get a six-band Hamiltonian $\Tilde{H}_6$, written as
\begin{equation}
    \Tilde{H}_6=\sigma_0\otimes H_3+H_{\text{SOC}},
    \label{intermédiaire spinful low-energy}
\end{equation}
where $H_{\text{SOC}}$ is expressed in the trigonal basis, and $H_3$ is given by Eq.(\ref{H3quad}). Further manipulation detailed in appendix \ref{annexe spinful low-energy}, using the experimentally relevant simplifying approximation $\lambda=3d$, yields the two-band Hamiltonian $H_2=h_0(\mathbf{k})\sigma_0+\mathbf{h}(\mathbf{k})\cdot\mathbf{\sigma}$, with
\begin{equation}
    h_0(\mathbf{k})=-2t-t_d+\big(1-3\sqrt{3}\big)d+\frac{2t+t_d}{4}k^2,
    \label{eq:lowenergy:h0}
\end{equation}
and
\begin{equation}
    \mathbf{h}(\mathbf{k})=\frac{c_0}{2}\Big(-2k_x,k_x+\sqrt{3}k_y,k_x-\sqrt{3}k_y\Big).
    \label{eq:lowenergy:h}
\end{equation}
Note that the spin-orbit coupling mixes all three bonding spinless bands, producing three Kramers pairs of energy branches separated by an energy gap at the $\Gamma$ point. The low-energy model focuses on the lowest Kramers pairs.
        \subsubsection{Dispersions}
\begin{figure}[ht!]
            \centering
            \includegraphics[width=\columnwidth]{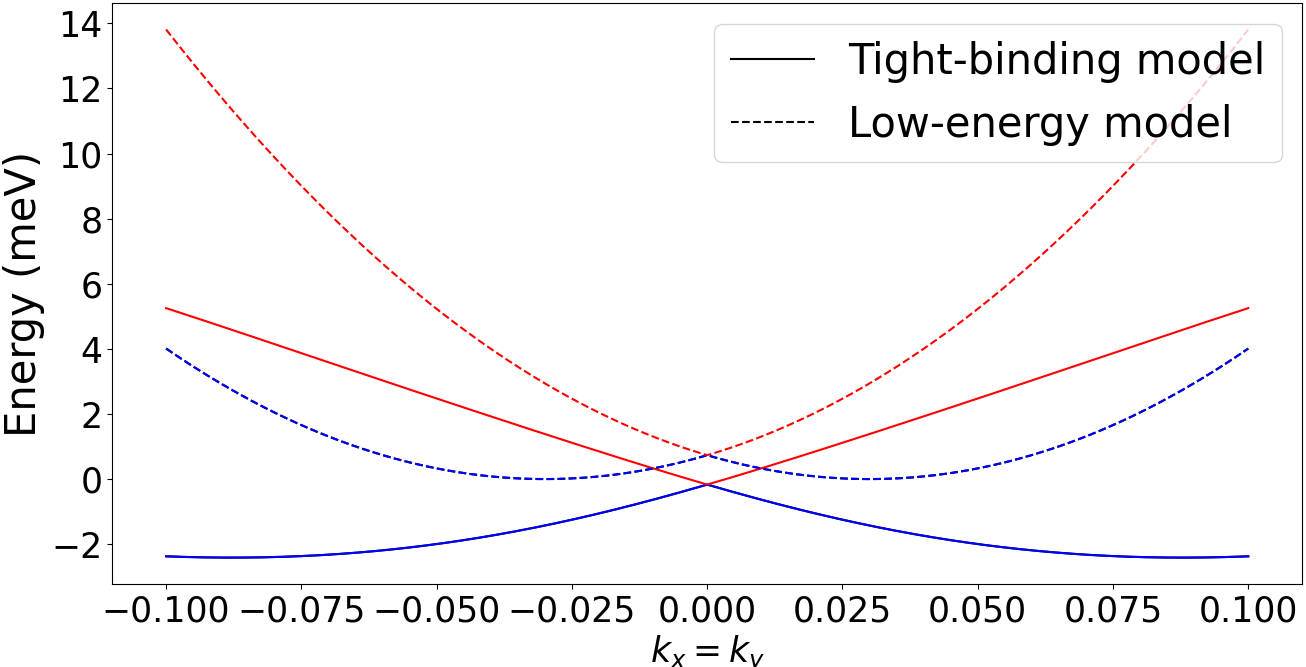}
            \caption{Energy dispersions of the low-energy two-band model $H_2$ overlaid with the two lowest band of the full TB model $H_{12}$, near $\Gamma$ and in the $k_x=k_y$ direction. The energy is defined with respect to the crossing point of the solid lines.}
            \label{dispersions_2bandes}
        \end{figure}

        The  dispersions are given by
        \begin{equation}
            \epsilon_\pm=-2t-t_d+\big(1-3\sqrt{3}\big)d\pm\sqrt{\frac{3}{2}}c_0k+\frac{2t+t_d}{4}k^2.
        \end{equation}
        As in the spinless case, we compare the dispersions of the latter with those of the full TB model $H_{12}$ in Fig. \ref{dispersions_2bandes}.
        \subsubsection{Quantum/band geometry}
    \paragraph{Quantum metric}
The quantum metric is the same for the two bands, being equal to \cite{liang_band_2017}
\begin{equation}
    g_{\mu\nu}=\frac{k^2\delta_{\mu\nu}-k_\mu k_\nu}{4k^4}.
\end{equation}
\paragraph{Berry curvature}
As for the Berry curvature, the upper and lower parts of the Dirac cone yield opposite $\delta(k)$ contributions at the contact point. Note that in the full model, there are two sources of Berry curvature. One is, as we just mentioned, the spin-orbit coupling, which results in the mentioned Dirac cone. The other is the mixing of the lowest band with the others which drives a Berry curvature the order of unity in $a_0^2$, as mentioned in Sec. \ref{sec:Poids-TB}.

    \subsection{Superfluid weight in the low-energy model}
\label{sec:qualitative SFW}
During the past decade, a significant number of papers have discussed the role of the normal-state quantum metric on the superconducting state \cite{liang_band_2017,torma_superconductivity_2022,rossi_quantum_2021}. It was found that the tensor relating the supercurrent to the electrodynamic perturbation of a superconductor produces two types of terms. One type is the well-known BCS contribution (see Ref. \cite{chandrasekhar_superconducting_1993}, for example), and the other is a geometric contribution which stems from interband couplings when the normal-state electronic structure involves several bands. In the isolated band limit, and in two-band models, this geometric contribution directly involves the normal-state quantum metric \cite{liang_band_2017}. 

Initially, the theory was developed for flat bands where the conventional contribution vanishes and the geometric contribution then dominates. While we do not have flat bands, the Dirac cone structure of our low-energy model gives a strong quantum metric near the $\Gamma$ point. It thus seems relevant to investigate whether the normal-state quantum metric produces a sizeable effect on the superconducting state through this geometric superfluid weight. In the following two sections, we discuss the two contributions in the context of our low-energy model. We begin with a qualitative discussion aimed at explaining generic scenarios for the Berezinskii-Kosterlitz-Thouless (BKT) temperature versus the gate voltage $V_g$.  For the superconducting state, we assume a conventional $s$-wave pairing, which can accommodate the disordered nature of oxide interfaces. As for the value of the $s$-wave gap, it has been measured to be $\Delta=40\mu$eV at optimal doping for the $(001)$ LAO/STO interface \cite{Richter2013}, and a similar value for the (111) interface was reported in Ref. \cite{groen2016}.

\subsubsection{BKT Temperature}
In addition to the superfluid weight (which has the dimension of an energy in 2D), we consider the associated Berezinskii-Kosterlitz-Thouless (BKT) temperature, using the (isotropic) Nelson-Kosterlitz criterion \cite{liang_band_2017},
\begin{equation}
    T_{\text{BKT}}=\frac{\pi}{8k_B}D(T_{\text{BKT}}),
    \label{Neslon-Kosterlitz}
\end{equation}
where $D(T)$ is the superfluid weight at temperature $T$. Note that we make use of the isotropic criterion because in the studied system the superfluid weight turns to obey $D_{\mu\nu}=D\delta_{\mu\nu}$, both in the low-energy and tight-binding models.
The BKT temperature $T_\text{BKT}$ is the temperature above which  vortex-antivortex pairs start to unbind and thus destroy superconductivity. It is smaller than the critical temperature 
calculated in the framework of a mean-field approach. For $T_{\text{BKT}}$ not too close to $T_c$ the mean-field critical temperature, we may approximate $D(T_{\text{BKT}})$ by $D(T=0)$. This defines a ``mean-field'' BKT temperature which is larger than the actual one and  may thus give an upper bound estimate,
\begin{equation}
    T_{\text{BKT}}=\frac{\pi}{8k_B}D(T=0).
    \label{BKT T zéro}
\end{equation}
\subsubsection{Conventional contribution}
\label{subsec:Conventionnel_qualitatif}
The conventional contribution to the superfluid weight, at $T=0$, is given by \cite{liang_band_2017,chandrasekhar_superconducting_1993}
\begin{equation}
    D_{\mu\nu,\text{conv}}=\int_{\mathscr{S}_{\text{occ}}(\mu)}\mathcal D_2 \mathbf{k} \frac{\Delta^2}{E^3}(\partial_\mu\epsilon)(\partial_\nu\epsilon), 
    \label{Poids_conv}
\end{equation}
where $\mathscr{S}_{\text{occ}}(\mu)$ denotes the set of occupied states in the BZ at the chemical potential $\mu$, and $E=\sqrt{\epsilon^2+\Delta^2}$. $\mathcal D_2 \mathbf{k}$ is the integration measure for the hexagonal BZ. For isotropic linearly dispersing bands, Eq. (\ref{Poids_conv}) gives $D_{\mu\nu,\text{conv}}=\delta_{\mu\nu}D_{\text{conv}}$ with $D_{\text{conv}}\propto\sqrt{\Delta^2+\mu^2}$ \cite{liang_band_2017}, where $\Delta$ is the s-wave superconducting gap. In appendix \ref{annexe calcul Dconv}, we show that a similar result holds for a general isotropic quadratic band. Consequently, the conventional superfluid weight is essentially proportional to the chemical potential $\mu$ for $\mu\gtrsim\Delta$. 
\subsubsection{Geometric contribution}
\label{subsec:Géométrique_qualitatif}
    \paragraph{General expression}
The geometric contribution at $T=0$ for a two-band Hamiltonian of the form $h_0\sigma_0+\mathbf{h}\cdot\bm{\sigma}$, is given by \cite{liang_band_2017}
\begin{equation}
    D^{\pm}_{\text{geom},\mu\nu}=\pm\int_{\mathscr{S}_{\text{occ}}(\mu)} \mathcal D_2 \mathbf{k}\frac{4\Delta^2}{E_\pm}\frac{h}{\mu-h_0}g_{\mu\nu},
    \label{Poids_geom}
\end{equation}
and more generally the geometric contribution is linked to the quantum metric in the so-called isolated limit \cite{liang_band_2017}, noting that Eq. (\ref{Poids_geom}) does not rely on it. Furthermore, note the additional factor of two with respect to the expression given in Ref. \cite{liang_band_2017}. This is because the definition of the metric in Ref. \cite{liang_band_2017} is twice the usual one \cite{berry_quantum_1989,graf_berry_2021}.
    \paragraph{Qualitative $\mu$ dependence}
Because of the Dirac cone at the $\Gamma$ point, the quantum metric of the two lowest bands in our model will exhibit a strong peak. Even in the spinless case, the lowest band is accompanied by a strong peak at the $\Gamma$ point. We therefore consider the case where the bands exhibit a peaked quantum metric around where the zero of the chemical potential is defined, hereafter called the zero-filling point. 

\begin{figure}[ht!]
    \centering
    \includegraphics[width=\columnwidth]{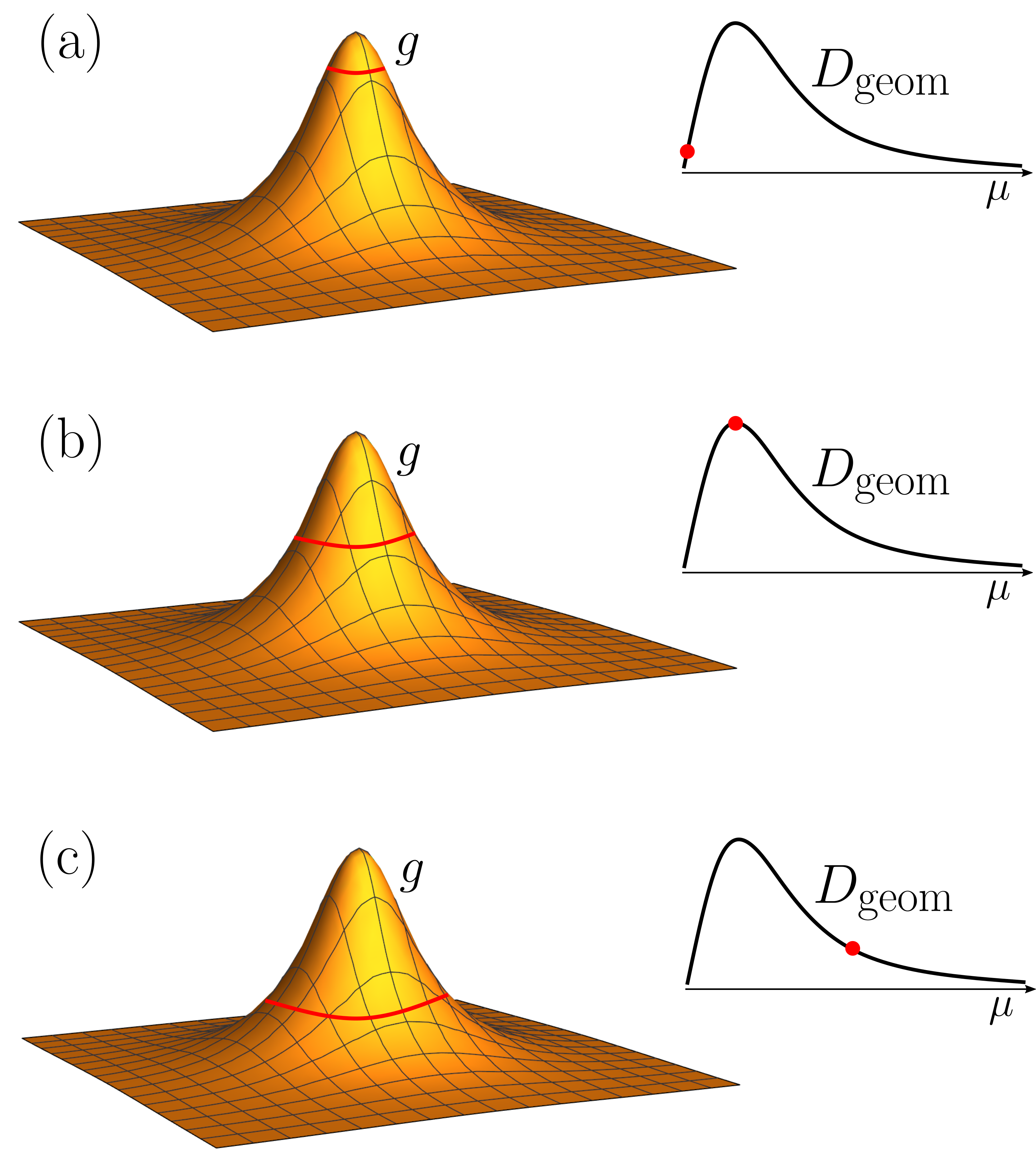}\caption{Emergence of a dome in the geometric superfluid weight from the Fermi contour contribution (in red) in Eq.(\ref{Poids_geom}) and its corresponding location in the dome (red dot). (a) Low-$\mu$ regim. (b)  Intermediate-$\mu$ regime. (c) High-$\mu$ regime}.
    \label{Pic_metrique}
\end{figure}

The $1/E(\mathbf{k})$ factor in Eq. (\ref{Poids_geom}) enhances the contribution at the Fermi contour, making it dominant. Focusing on this contribution, we can propose a scenario explaining the emergence of a dome in the geometric superfluid weight when the metric has a peak at the zero-filling point. We sketch this scenario in Fig. \ref{Pic_metrique}.

 At low $\mu$, the band starts to be filled around $\Gamma$. The Fermi contour is thus at the top of the peak, but it is also narrow, such that $D_{\text{geom}}$ is low. However, as the filling increases, the Fermi contour gets wider while still being high and thus $D_{\text{geom}}$ becomes larger. This is the low-$\mu$ regime, shown in Fig. \ref{Pic_metrique}a. 
 
 The chemical potential $\mu$ then reaches a value where the trade-off between the height and the extent of the Fermi contour is optimal, and $D_{\text{geom}}$ reaches its maximal value. This is the intermediate-$\mu$ regime in Fig. \ref{Pic_metrique}b. 
 
 Beyond the latter, the Fermi contour still gets wider but not enough to compensate the smaller values of $g_{\mu\nu}$, resulting in a decrease of $D_{\text{geom}}$. This is the high-$\mu$ regime in Fig. \ref{Pic_metrique}c.
\subsubsection{Qualitative $\mu$-dependence of the BKT temperature}
\begin{figure}[ht!]
    \centering
    \includegraphics[width=\columnwidth]{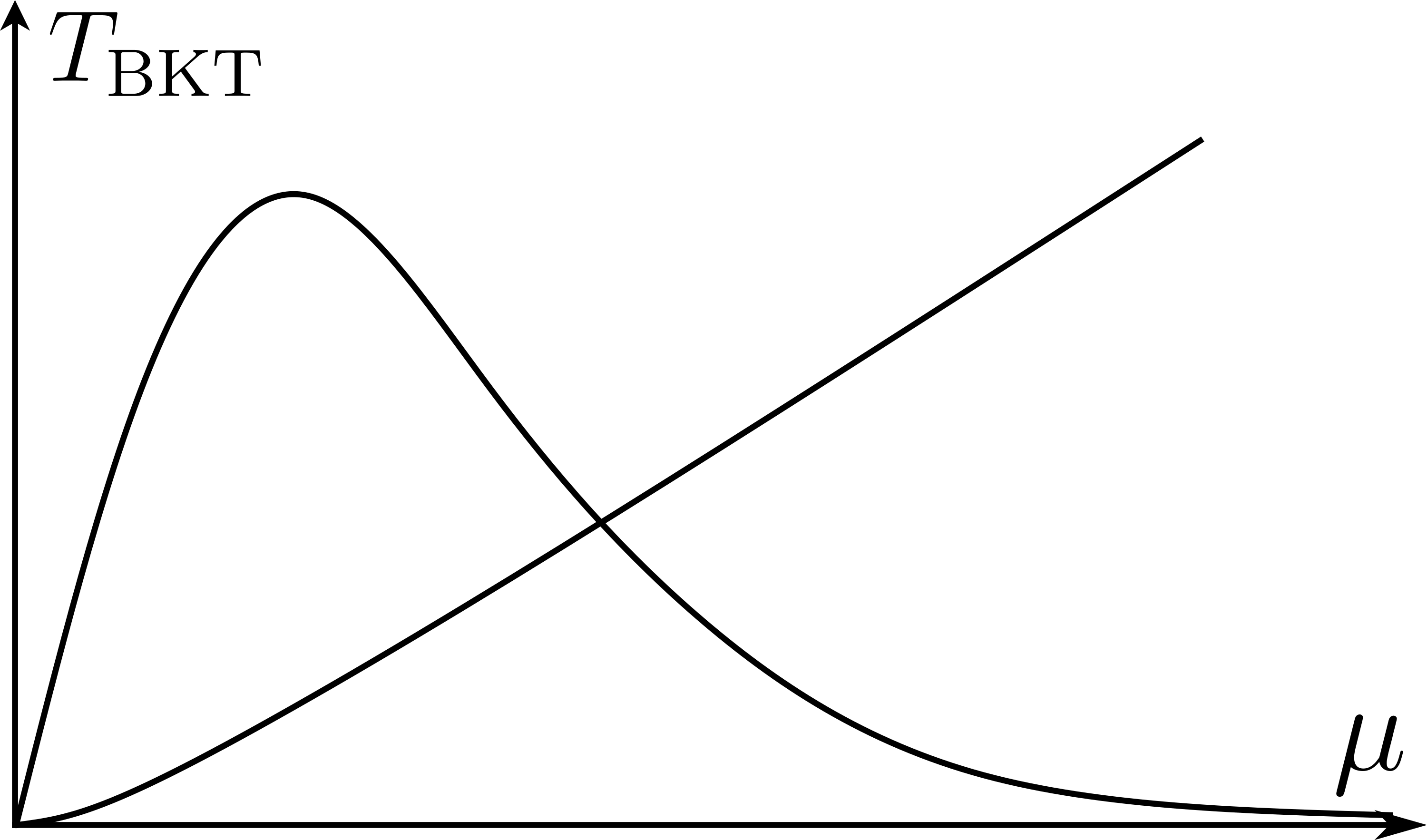}
    \caption{Qualitative dependence of the geometric contribution, with a dome behavior, and the conventional contribution, with a linear behavior, to the BKT temperature of Eq.(\ref{BKT T zéro}).}
    \label{BKT_qualitatif}
\end{figure}

Following the above discussion, we sketch the qualitative evolution of the BKT temperature given by Eq. (\ref{BKT T zéro}) as shown in 
Fig. \ref{BKT_qualitatif}.

In order to showcase the role of the geometric part,  Fig. \ref{BKT_qualitatif} is a plot of the conventional and geometric contributions in a situation when both quantities are of the same order. We show below that this is the case for the low-energy and full TB model.

\section{ Relating the superfluid weight to the experimental data}
\label{sec:semiquantitative SFW}
    \subsection{Thermal and disorder effects}
     Experimental studies of the superconducting transition in (001) and (111) oriented LAO/STO interfaces indicate that the BKT scenario is indeed relevant \cite{groen2016, Monteiro2017, Manca2019, Lesne2021}.  The superfluid weight is proportional to the superfluid carrier density $n_s$. Temperature effects could be included through the temperature dependence of the SC gap and the Fermi factor \cite{Tinkham1995}, however disorder effects would also need to be properly included, a task beyond the scope of the present work. Microwave measurements of the London penetration depth by Lesne \textit{et al.} \cite{Lesne2021} show that the value of $n_s$ at  $T_{BKT}$ is one order of magnitude smaller than it is at "zero temperature". These experiments contain both disorder and thermal effects. Therefore, one way to take these factors into account would be applying such a renormalisation to the superfluid density and thus to the superfluid weight. While we do not provide a full-fledged theory of the origin of this reduction of the superfluid density, we take it as an experimental fact, and hereafter we qualitatively renormalize both the conventional and geometric contributions to the superfluid weight by a factor $1/10$. 
   
    \subsection{Superfluid weight from the low-energy model}

\begin{figure}[ht!]
        \centering
        \includegraphics[width=\columnwidth]{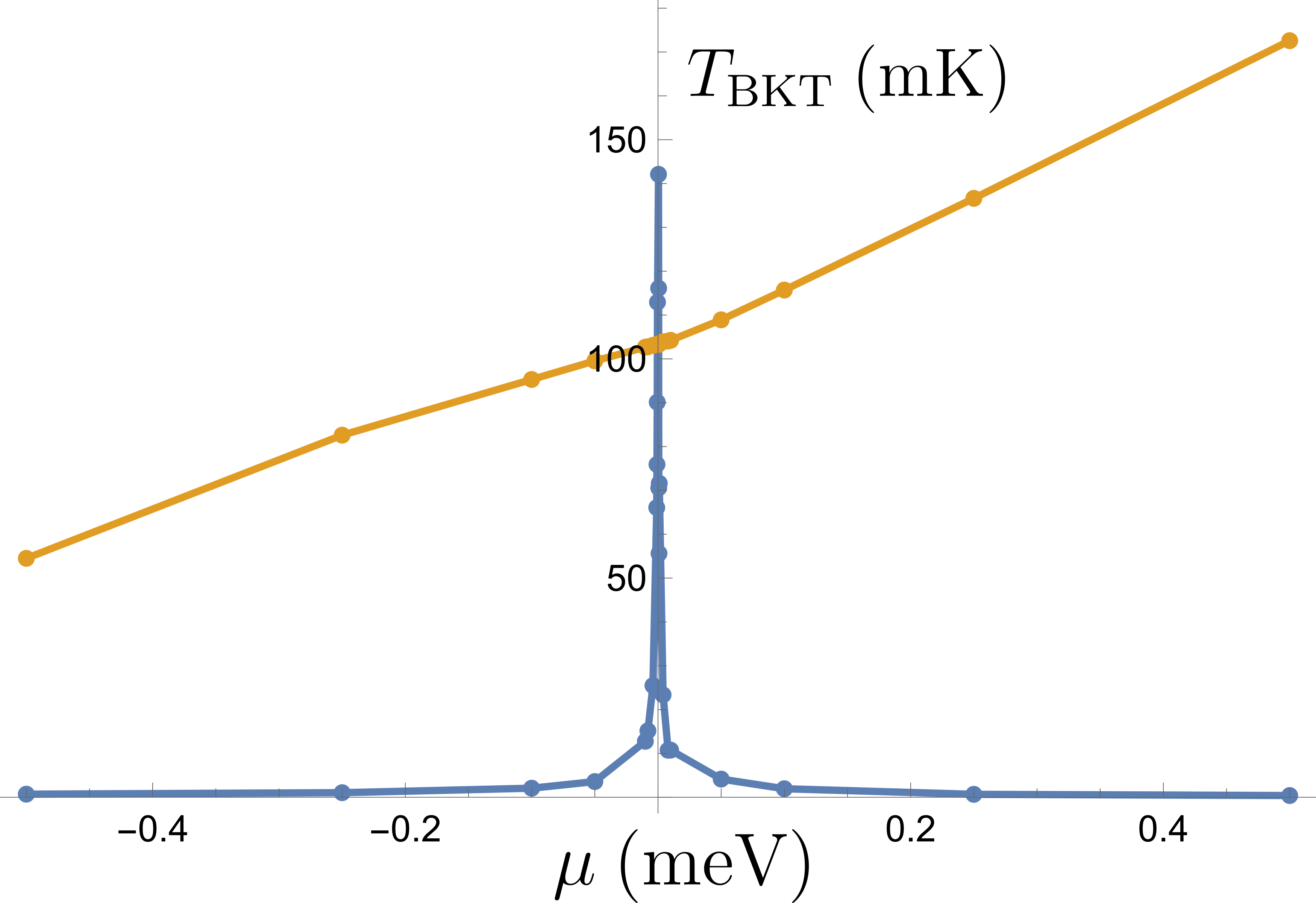}
        \caption{Conventional (orange) and geometric (blue) contributions to the superfluid weight from the low-energy model of Eqs.(\ref{eq:lowenergy:h0},\ref{eq:lowenergy:h}) using Eq.(\ref{Poids_geom}). The chemical potential $\mu$ is defined with respect to the Dirac point of the low-energy model. $T_{\text{BKT}}$ is defined in Eq.(\ref{BKT T zéro}).}
        \label{fig:Poids_lowenergy}
    \end{figure}

     We now numerically compute both contributions of the superfluid weight for the low-energy model, using Eq. (\ref{Poids_geom}). The result is plotted in Fig. \ref{fig:Poids_lowenergy}, and is indeed consistent with the analysis made in Sec.\ref{sec:qualitative SFW}. Note that the chemical potential is defined with respect to the Dirac point of the low-energy model.

    The conventional contribution is essentially linear in the chemical potential $\mu$. Note that it is non-zero below the Dirac point because the minimum of the lower band is located away from the $\Gamma$ point, as seen in
    Fig.~\ref{dispersions_2bandes}. Notice the small dip seen near $\mu=0$ in Fig.~\ref{fig:Poids_lowenergy}, which might be due to a reduction of the density of states, because of the Dirac point.
    
    The evolution of the geometric contribution is consistent with the scenario depicted in Sec.\ref{sec:qualitative SFW} and
    Fig.~\ref{Pic_metrique}. Because of the Dirac point, the quantum metric exhibits a strong (in fact divergent) peak at $\Gamma$, i.e. $\mu=0$ in Fig. \ref{fig:Poids_lowenergy}. For $\mu<0$, the Fermi contour is at the bottom of the metric's peak, so the geometric contribution is low. But then as $\mu$ gets closer to zero, the Fermi contour moves to a higher position on the metric's peak. The metric diverges at the degeneracy point so that the geometric contribution  diverges at $\mu=0$. For $\mu>0$, the Fermi contour gradually moves to the bottom of the peak, and the geometric contribution goes back to zero. All this happens on a scale of 1meV around the Dirac point, as seen in
    Fig. \ref{fig:Poids_lowenergy}. We note that the variation of the geometric contribution is so steep around $\mu=0$ that one would need to get to very low values of $\mu$ in order to reach the regime where the geometric contribution dominates. 
    
     \subsection{Superfluid weight from the tight-binding model}
     \label{sec:Poids-TB}
      
    \begin{figure}[ht!]
        \centering
        \includegraphics[width=\columnwidth]{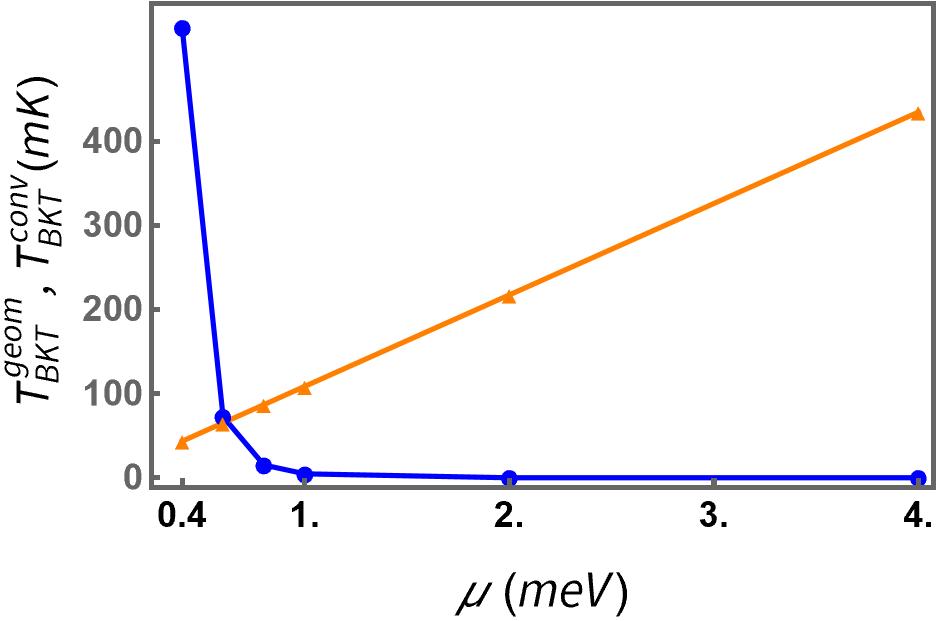}
        \caption{Conventional (orange) and geometric (blue) contributions to the superfluid weight from the spinful TB model. The chemical potential $\mu$ is defined with respect to the Dirac point of the spinful TB model.}
        \label{fig:poids_TB}
    \end{figure}
    As a test of the accuracy of the low-energy model, we can also compute the superfluid weight from the spinful TB model, as shown in Fig. \ref{fig:poids_TB}. The chemical potential $\mu$ is defined with respect to the Dirac point of the spinful TB model. Note that although one would \textit{a priori} have to use the multi-band formulas given in Ref.\cite{liang_band_2017}, we chose to use the two-band formula given in Eq.(\ref{Poids_geom}) in the following way. We took the band dispersions of the full TB model and considered their sum and differences, whose halves give an effective $h_0$ and $h$, respectively. We then plug in those along with the quantum metric of the lowest band, coming again from the full TB model, in the two-band formula.
    We see that our qualitative arguments of section \ref{sec:qualitative SFW} hold here as well. The geometric contribution shows a divergent behaviour in the $\mu\ll1$meV regime, and precipitously drops to zero when we increase $\mu$. The conventional contribution is linear in the chemical potential. One difference between 
    Fig. \ref{fig:poids_TB} and the superfluid weight of the low-energy model is that the two contributions become equal for a larger value of $\mu$. This may be due to the fact that bands in both the anti-bonding group, despite being $7$eV higher in energy, and the other bonding bands have a non-negligible overlap with the two bands of interest. Note that this is also the case for the Berry curvature, where this overlap between the lowest and the highest bands generates a Berry curvature of order $a_0^2$.
    
\subsection{Gate voltage dependence: Two-dome scenario}

\begin{widetext}

    \begin{figure}[ht!]
        \centering
        \includegraphics[width=\textwidth]{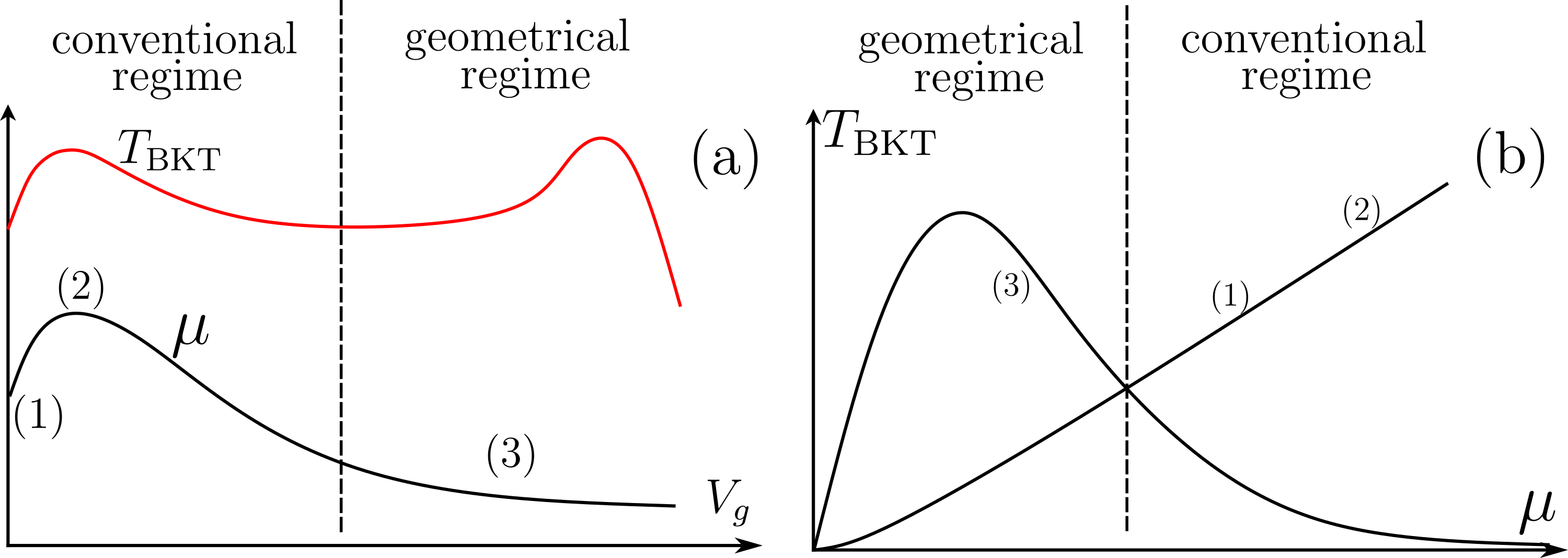}
        \caption{(a) Proposed dependence of the chemical potential on the gate voltage, with the resulting BKT temperature in red. (b) Correspondence between the qualitative $\mu$ and $V_g$ dependencies of the geometric and conventional contributions. In both (a) and (b) the areas where the geometrical and conventional contributions dominate are outlined. The vertical dashed line in (a) and (b) marks the boundary between the two regimes.}
        \label{Correspondance_mu_Vg}
    \end{figure}
    
\end{widetext}

We have discussed the qualitative dependence of the BKT temperature [Eq.(\ref{BKT T zéro})] on the chemical potential. However, the experimentally observed superconducting dome \cite{rout_link_2017,bert_gate-tuned_2012,gariglio_interface_2015,Gariglio2016, Monteiro2019} is measured when one tunes the gate voltage $V_g$ or the conductivity of the interface. There are strong indications \cite{Monteiro2019,Khanna2019} suggesting that  the (Hall) carrier density (or the chemical potential) has a non-monotonic dependence on $V_g$ or on the conductivity. In the following parts, we take the characteristic dependence of the chemical potential on the gate voltage, which is depicted in Fig. \ref{Correspondance_mu_Vg}a, as an empirical input and do not aim at a microscopic understanding of its origin that may be due to electronic correlations or complicated charge transfer processes between the interface and the bulk material. Consequently, there is no simple correspondence between the superconducting domes that result from changing $\mu$ and the phase diagram that one obtains upon changing $V_g$. We therefore emphasize that we discuss a qualitative scenario here, based on available experimental data, and that a knowledge of the quantitative dependence of the chemical potential on the gate voltage is required in order to get quantitative agreement with experiments.

The scenario depicted in Fig. \ref{Correspondance_mu_Vg} may be understood as follows. At zero gate voltage ($V_g=0$), the system is slightly doped and has a non-zero electronic density [point (1) in Fig. \ref{Correspondance_mu_Vg}a]. In this density range, we hypothesize that the superfluid weight is dominated by the conventional contribution  (point (1) in Fig. \ref{Correspondance_mu_Vg}b). Upon turning on the (positive) gate voltage, the chemical potential first increases up to point (2), as one expects from the usual linear dependence of $\mu$ on $V_g$. Consequently, the BKT temperature increases (see Fig. \ref{Correspondance_mu_Vg}b) in what may be called the \textit{underdoped regime}. However, beyond point (2), we are confronted with the above-mentioned non-monotonic dependence of the chemical potential, which starts to \textit{decrease} upon further increasing $V_g$, and so does the BKT temperature (\textit{overdoped regime}). Eventually the density is depleted to the extent that the chemical potential is situated in a regime dominated by the geometric contribution to the superfluid weight, which thus saturates. This crossover is represented by the vertical dashed line in Fig. \ref{Correspondance_mu_Vg}. The experimentally observed dome in the BKT temperature is therefore driven by the non-monotonic dependence of $\mu$ on $V_g$ in a regime dominated by the conventional superfluid weight.

 As mentioned, the above scenario is based on an empirically found non-monotonic behavior of the chemical potential (the electronic density measured in Hall experiments) on the gate voltage \cite{Monteiro2019,Khanna2019}. This scenario has very recently found further experimental support (see Ref. \cite{Sala2024}) and it might find a further corroboration in experiments at higher gate voltages. In this case, according to our scenario, the density might become so low that one enters a regime that is dominated by the geometric contribution to the superfluid weight, in which case one may expect another increase of the BKT temperature, as depicted by point (3) in Fig. \ref{Correspondance_mu_Vg}. Eventually, there may then exist even a second dome that is entirely due to the geometric superfluid weight in the low-density limit, if the chemical potential keeps decreasing. In summary, the measured domes would be caused by the non-monotonic variation of the chemical potential with respect to the gate voltage such that there should be a \textit{secondary superconducting dome},  coming from the geometric contribution, for higher values of the gate voltage. The evolution of the critical temperature (or superfluid density, BKT temperature) would be similar to that sketched in Fig. \ref{Correspondance_mu_Vg}a (red curve), as long as only the lowest energy band contributes to the superfluid condensate. 

According to our picture, the two superconducting domes that one expects upon increasing the gate voltage have thus different origins. The first one corresponds to the regime when the conventional superfluid weight dominates and the second one to the regime when the geometric weight dominates.

\section{Conclusion}

Our study underscores the impact of the normal state quantum geometry on the superconducting state of the (111) $\text{LaAlO}_3/\text{SrTiO}_3$ interface. 
Building upon a tight-binding modeling of the interface, we developed a two-band low-energy model around the $\Gamma$ point. This low-energy model reveals an isotropic Dirac cone at the $\Gamma$ point, driven by the orbital mixing. From the low-energy model, we then drew a qualitative scenario for the chemical potential ($\mu$) dependence of the conventional and geometric contributions to the superfluid weight. The conventional contribution is suggested to be linear in $\mu$, because of the dispersion's isotropy. We also argued that the strong peak in the quantum metric coming from the Dirac cone results in a dome-shaped behavior of the geometric contribution  as a function of $\mu$. 

Finally, we probed the relevance of our scenario to experiments in the case of the LAO/STO interface. We first effectively took into account thermal and disorder effects by renormalizing the superfluid weight. Then, we numerically computed the superfluid weight from the two-band low-energy model 
and the spinful TB model. In both cases, we find the right order of magnitude for the superfluid weight (and the associated BKT temperature). The most significant difference between the two models is the value of $\mu$ below which the geometric contribution becomes dominant. In the low-energy model, the geometric contribution exhibits a very narrow peak around $\mu = 0$ and therefore the regime where it dominates would seem hard to observe, in practice. In the spinful TB model, bands from the anti-bonding group, despite being $6-7$ eVs higher up, and other bonding bands, contribute significantly to the quantum metric of the lowest Kramers’ partners, thereby making the geometric contribution less
steep and increasing the value of chemical potential below which the geometric contribution dominates. We then argue that, despite its simplicity, our low-energy model captures the evolution of the superfluid weight reasonably well.

Experiments tune the 2DEG with a gate voltage $V_g$, so that we aimed to qualitatively describe the behaviour of the superfluid weight as a function of said $V_g$. Hall transport experiments \cite{Monteiro2019,Khanna2019} find a non-monotonic dependence of the Hall carrier density on $V_g$. Assuming that the Hall carrier density is monotonic in the chemical potential, this means that the chemical potential is non-monotonic in $V_g$. Taking this as an experimental fact, we then inferred a qualitative dependence of the superfluid weight (and thus BKT temperature) on the gate voltage. The experimentally observed dome is suggested to come from the conventional contribution and the non-monotonic dependence of $\mu$ on $V_g$. Extrapolating further, we also suggest the saturation of the dome mentioned earlier and the appearance of a \textit{second superconducting dome}, due purely to the geometric contribution to the superfluid weight.

Given the ubiquituousness of quantum geometry, this \textit{hidden influence} on the superconducting state might be apparent in other classes of materials. Finally, this positive effect of the normal-state quantum metric on superconductivity needs to be contrasted with a previous theoretical discussion \cite{simon_role_2022} suggesting a negative impact of the normal-state Berry curvature on superconductivity. This could suggest a \textit{normal state curvature-metric competition} towards superconductivity.

\section*{Acknowledgements}
  We wish to acknowledge Frédéric Piéchon for his insightful input on our work and careful reading of our manuscript. We thank Andrea Caviglia and Roberta Citro for valuable discussions.


\bibliographystyle{apsrev4-2}
\bibliography{Bibliolaosto_r4}

\begin{appendix}
\section{Bilayer model}\label{Annexe_bilayer}
\begin{widetext}

    \begin{figure}[ht!]
        \includegraphics[width=0.8\textwidth]{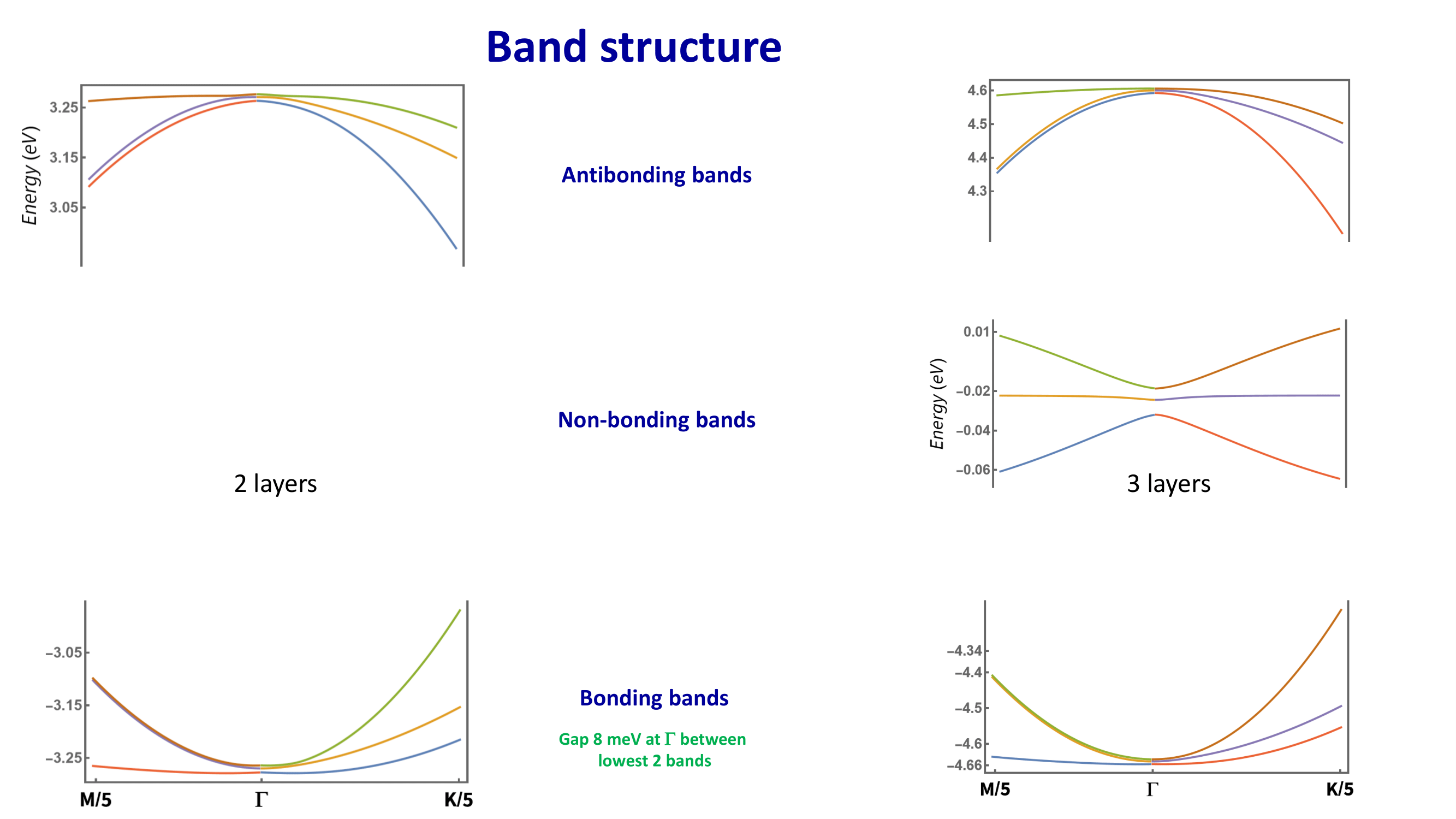}
        \caption{Comparison between the bilayer (left) and trilayer (right) TB models.}
        \label{Compare23couches}
    \end{figure}
    
\end{widetext}

Here, we explain in more detail the choice shown in Fig. \ref{Schema_LAOSTO} to only include two layers of Ti atoms in the TB model. As said in the main text, one should \textit{a priori} consider three layers, so a nine band TB model (without spin). In Fig. \ref{Compare23couches} we show a comparison of the band dispersions obtained from the bilayer TB model (on the left) and a similarly obtained nine band trilayer TB model.
First, we see that the bonding and anti-bonding triplets are not visibly modified by the addition of the third layer. Furthermore, the three additional bands are approximately at zero energy, and therefore several eVs away from the band we are concerned about, in the bonding triplet. Second, also the eigenstates of the bonding triplet should not be modified by the additional bands because, again, of their energy difference. Finally, from Eq.(\ref{QGTHam}), we also see that following the same argument, the quantum geometric tensor \cite{berry_quantum_1989} should not be modified by the additional bands.
\begin{equation}
    Q^n_{\mu\nu}=\sum_{m\neq n}\frac{\bra{u_n}\partial_\mu H\ket{u_m}\bra{u_m}\partial_\nu H\ket{u_n}}{(E_n-E_m)^2}.
    \label{QGTHam}
\end{equation} 
In conclusion, we can reasonably discard the third layer and consider a bilayer model of Ti atoms, as shown in Fig. \ref{Schema_LAOSTO}.
\section{Expression of the kinetic terms}
\label{expressions cinetiques}
\begin{figure}[ht!]
        \centering
        \includegraphics[width=\columnwidth]{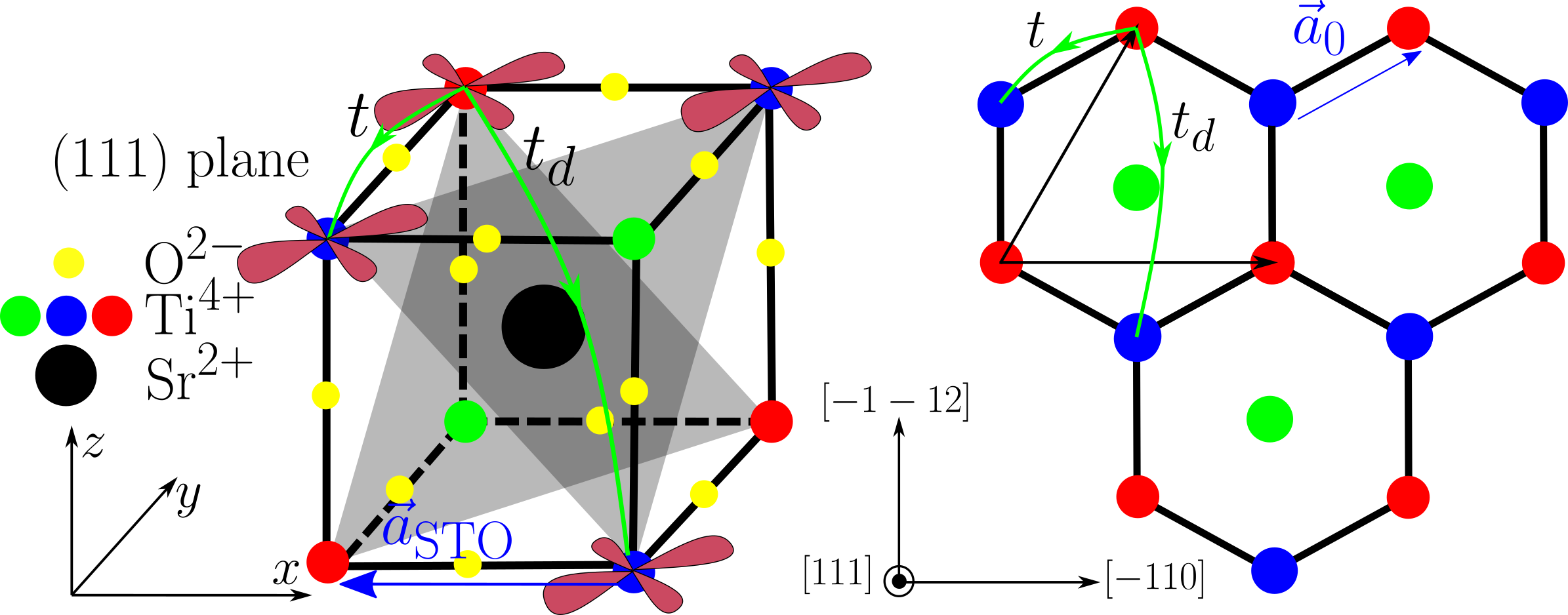}
        \caption{Diagram of the inter-layer and intra-orbital hoppings for the $d_{xy}$ orbitals. On the left, the 3D cubic lattice and the $d_{xy}$ orbitals of the Ti atoms. Hopping paths and corresponding amplitudes are shown as green color lines. On the right, the associated projection in the $(111)$ orientation.}
        \label{orbitales_laosto}
\end{figure}

Hoppings are between two neighboring layers, with amplitude $t$ for $\pi$-hoppings and $t_d$ for $\delta$-hoppings between blue and red sites (Fig. 1). The origin of the basis lattice vectors is chosen at the center of an hexagon. $e$, $f$ and $g$ have the following expressions \cite{xiao_interface_2011,rodel_orientational_2014},
\begin{widetext}
\begin{equation}
    \begin{cases}
        e=-\bigg\{\exp(ik_y)+\exp\bigg[i\Big(\frac{\sqrt{3}}{2}k_x-\frac{1}{2}k_y\Big)\bigg]\bigg\}-r\exp\bigg[-i\Big(\frac{\sqrt{3}}{2}k_x+\frac{1}{2}k_y\Big)\bigg]\\
        f=-\bigg\{\exp(ik_y)+\exp\bigg[-i\Big(\frac{\sqrt{3}}{2}k_x+\frac{1}{2}k_y\Big)\bigg]\bigg\}-r\exp\bigg[i\Big(\frac{\sqrt{3}}{2}k_x-\frac{1}{2}k_y\Big)\bigg]\\
        g=-\exp\Big(-\frac{i}{2}k_y\Big)\times2\cos\Big(\frac{\sqrt{3}}{2}k_x\Big)-r\exp(ik_y)
    \end{cases}
    \hspace{2mm}\text{with}\hspace{2mm}
    r=\frac{t_d}{t}.
\end{equation}
\end{widetext}
As example, we schematize the case of the $d_{xy}$ orbitals corresponding to the hopping term $g$ in Fig. \ref{orbitales_laosto}, and we derive $g$.
 First, consider the hopping pictured on the right of Fig. \ref{orbitales_laosto}, going from the top red atom to the neighbouring left blue atom, with an amplitude of $t$. As seen in Fig. \ref{orbitales_laosto}, the associated overlap in the cubic lattice is a $\pi$-overlap through the intermediate O site. The amplitude of this hopping is $t=1.6$eV. The associated phase is $\text{exp}(-i\mathbf{k}\cdot\mathbf{a}_2)$ with $\mathbf{a}_1=(-\sqrt{3}/2,3/2)$, by equivalence with the red atom in the same motif. The similar hopping to the right also has amplitude $t$ and is associated to the phase $\text{exp}(-i\mathbf{k}\cdot\mathbf{a}_1)$ with $\mathbf{a}_1=(\sqrt{3}/2,3/2)$. The other hopping from the top red atom to the lower blue atom is, as seen on the left of Fig. \ref{orbitales_laosto}, the result of the direct overlap of the $d$ orbitals of the two atoms. It is therefore a $\delta$-hopping, associated with an energy amplitude $t_d=70$meV. As for its phase, it does not come with a phase shift, as the two atoms are part of the same motif. This way, and shifting every term by $\text{exp}(ik_y)$ so that the origin is at the central green atom, the interlayer $d_{xy}-d_{xy}$ hopping term is given by
 
\begin{align}
    g&=-te^{-i/2(-\sqrt{3}k_x+k_y)}-te^{-i/2(\sqrt{3}k_x+k_y)}-t_de^{ik_y}\nonumber\\
    &=-2te^{-ik_y/2}\cos\bigg(\frac{\sqrt{3}}{2}k_x\bigg)-t_de^{ik_y}.
\end{align}

\section{Derivation of the orbital mixing term}
\label{annexe OM}

In the orbital basis, $H_{\text{om}}$ can be written as $A\otimes B$ with $A$ and $B$ acting in the layer and orbital subspaces respectively. The orbital mixing term consists of the interlayer and interorbital couplings so the diagonal elements of $A$ and $B$ must vanish. The hexagonal lattice structure seen in Fig. \ref{Schema_LAOSTO} has the $\mathcal{C}_{3v}$ symmetry. In order to respect the latter, we assume that all the couplings have the same magnitude in energy and that those between layer 1 to layer 2 are the same as those from layer 2 to layer 1. Therefore we can write 
\begin{equation}
    A\otimes B=c_0\begin{pmatrix}
        0&a\\
        a&0
    \end{pmatrix}
    \otimes
    \begin{pmatrix}
        0&b_1&b_2\\
        b_3&0&b_4\\
        b_5&b_6&0
    \end{pmatrix},
\end{equation}
where the coefficients are complex, and of modulus 1 in order to have the same magnitude $c_0$ in energy. They are further constrained by the fact that the term must be Hermitian. Using $(A\otimes B)^\dagger=A^\dagger\otimes B^\dagger$, this means that we must have $a^*=a$, $b_3=b_1^*$, $b_5=b_2^*$ and $b_6=b_3^*$. We choose $a=1$, such that
\begin{equation}
    A\otimes B=c_0\tau_x\otimes\begin{pmatrix}
        0&b_1&b_2\\
        b_1^*&0&b_3\\
        b_2^*&b_3^*&0
    \end{pmatrix}.
\end{equation}
We then introduce $(\phi_i,\psi_i)$ such that $b_i(\mathbf{k})=\operatorname{cos}(\phi_i(\mathbf{k}))+i\operatorname{sin}(\psi_i(\mathbf{k}))$. We look for $\phi_i$ and $\psi_i$ that are linear combinations of $k_x$ and $k_y$, which is natural for TB models. These hoppings are also antisymmetric under an inversion operation $\mathbf{r}\longmapsto-\mathbf{r}$ \cite{khalsa_theory_2013}, which adds the constraint $b_i(-\mathbf{k})=-b_i(\mathbf{k})$ since in $\operatorname{exp}(i\mathbf{k}\cdot\mathbf{r})$ doing $\mathbf{r}\longmapsto-\mathbf{r}$ is equivalent to $\mathbf{k}\longmapsto-\mathbf{k}$. Hence, $\operatorname{cos}(\phi_i(\mathbf{k}))=0$. Writing the allowed hoppings between the red and blue sites explicitly (Fig. \ref{Schema_LAOSTO}) with the above requirements then gives $b_i(\mathbf{k})=i\operatorname{sin}(\psi_i(\mathbf{k}))=i\operatorname{sin}(\alpha_ik_x+\beta_ik_y)$, with $\psi_i(\mathbf{k})=\alpha_ik_x+\beta_ik_y$. Next, the orbital mixing term needs to obey the $\mathcal{C}_{3v}$ symmetry, i.e. a $2\pi/3$ rotation with an axis perpendicular to the $(111)$ plane and a mirror symmetry parallel to the $(\overline{11}2)$ orientation. The $2\pi/3$ rotation transforms $\mathbf{r}=(x,y,z)$ into $\mathbf{r}'=(z,x,y)$ in the original cubic unit cell. Therefore the orbitals are transformed as $(d_{yz},d_{xz},d_{xy})\longmapsto(d_{xy},d_{yz},d_{xz})$. In order to obey this $\mathcal{C}_3$ symmetry, we must therefore have $b_1(\mathbf{k}')=b_3(\mathbf{k})$, $b_2(\mathbf{k}')=b_1^*(\mathbf{k})$ and $b_3(\mathbf{k}')=b_2^*(\mathbf{k})$ with $\mathbf{k}'=(-1/2k_x-\sqrt{3}/2k_y,\sqrt{3}/2k_x-1/2k_y)$. The mirror operation maps $\mathbf{r}=(x,y,z)$ to $\mathbf{r}'=(y,x,z)$ and $\mathbf{k}=(k_x,k_y)$ to $\mathbf{k}'=(-k_x,k_y)$, so that the orbitals transform as $(d_{yz},d_{xz},d_{xy})\longmapsto(d_{xz},d_{yz},d_{xy})$. In order to obey this symmetry, we must have $b_1(\mathbf{k}')=b_1^*(\mathbf{k})$, $b_2(\mathbf{k'})=b_3(\mathbf{k})$ and $b_3(\mathbf{k}')=b_2(\mathbf{k})$. These constraints on the $b_i$s put constraints on the coefficients $(\alpha_i,\beta_i)$ by taking the low-$k$ limit and identifying the $k_x$ and $k_y$ components (this is allowed since the constraints must be valid for all $\mathbf{k}$). The resulting system of equations puts five independent constraints such that $\beta_1=0$, $(\alpha_2,\beta_2)=(1/2\alpha_1,\sqrt{3}/2\alpha_1)$ and $(\alpha_3,\beta_3)=(-1/2\alpha_1,\sqrt{3}/2\alpha_1)$. We then recover Eq. (\ref{orbital mixing matrix}) with $\alpha_1=-\sqrt{3}$. 
\section{Gell-Mann matrices}
\label{annexe:Gell-Mann matrices}
\begin{align}
    &\Lambda_1=\begin{pmatrix}
        0&1&0\\
        1&0&0\\
        0&0&0
    \end{pmatrix},\hspace{1mm}\Lambda_2=\begin{pmatrix}
        0&-i&0\\
        i&0&0\\
        0&0&0
    \end{pmatrix}\nonumber\\
    &\Lambda_3=\begin{pmatrix}
        1&0&0\\
        0&-1&0\\
        0&0&0
    \end{pmatrix},\hspace{1mm}\Lambda_4=\begin{pmatrix}
        0&0&1\\
        0&0&0\\
        1&0&0
    \end{pmatrix}\nonumber\\
    &\Lambda_5=\begin{pmatrix}
        0&0&-i\\
        0&0&0\\
        i&0&0
    \end{pmatrix},\hspace{1mm}\Lambda_6=\begin{pmatrix}
        0&0&0\\
        0&0&1\\
        0&1&0
    \end{pmatrix}\nonumber\\
    &\Lambda_7=\begin{pmatrix}
        0&0&0\\
        0&0&-i\\
        0&i&0
    \end{pmatrix},\hspace{1mm}\Lambda_8=\frac{1}{\sqrt{3}}\begin{pmatrix}
      1&0&0\\
      0&1&0\\
      0&0&-2
    \end{pmatrix}.
\end{align}
\section{Trigonal basis}
\label{trigonal basis}
Let $U$ be the following unitary transformation in the $\{1,2\}\otimes\big\{\text{X},\text{Y},\text{Z}\big\}$ basis
\begin{equation}
        U=\tau_0\otimes P,\quad P=\begin{pmatrix}
        -\frac{1}{\sqrt{2}}&-\frac{1}{\sqrt{6}}&\frac{1}{\sqrt{3
        }}\\
        \frac{1}{\sqrt{2}}&-\frac{1}{\sqrt{6}}&\frac{1}{\sqrt{3}}\\
        0&\frac{2}{\sqrt{6}}&\frac{1}{\sqrt{3}}
    \end{pmatrix}.
\end{equation}
In this basis, the trigonal crystal field becomes diagonal,
\begin{equation}
    P^\dagger dH_d P=\begin{pmatrix}
        d&0&0\\
        0&d&0\\
        0&0&-2d
    \end{pmatrix}.
\end{equation}
The first two-fold degenerate eigenvalues represent the $e_{+g}$ and $e_{-g}$ orbitals while the third one represents the $a_{1g}$ orbital. They are separated by an energy gap of $3d\simeq 10$ meV. The kinetic term is transformed as
\begin{equation}
    H_{\text{cin}}=t\begin{pmatrix}
    \frac{e+f}{2}&\frac{e-f}{2\sqrt{3}}&-\frac{e-f}{\sqrt{6}}\\
        \frac{e-f}{2\sqrt{3}}&\frac{e+f+4g}{6}&-\frac{e+f-2g}{3\sqrt{2}}\\
        -\frac{e-f}{\sqrt{6}}&-\frac{e+f-2g}{3\sqrt{2}}&\frac{e+f+g}{3}
    \end{pmatrix},
\end{equation}
while the orbital mixing term becomes
\begin{equation}
    H_{\text{om}}=c_0\begin{pmatrix}
        0&iD&-iA\\
        -iD&0&iB\\
        iA&-iB&0
    \end{pmatrix},
\end{equation}
with
\begin{equation}
    \begin{cases}
        A=-\frac{1}{\sqrt{6}}\big(\alpha+\beta-2\delta\big)\\
        B=\frac{1}{\sqrt{2}}\big(\alpha-\beta\big)\\
        D=\frac{1}{\sqrt{3}}\big(\alpha+\beta+\delta\big).
    \end{cases}
\end{equation}

\section{Validity of the quadratic three-band approximation}
\label{validity annexe}
We now discuss the validity of the three-band approximation. Doing so amounts to neglecting the off-diagonal blocks in Eq. (\ref{H transforme}) which contain the confinement energy and the imaginary part of the kinetic term. From Ref. \cite{nakatsukasa_off-diagonal_2017}, the effect of such off-diagonal terms is in $\mathcal{O}\Big(\frac{||E||^2}{\text{gap}}\Big)$, where $E$ is the off-diagonal perturbation. The numerically observed gap (with our choice of parameters) is around $6-7$ eV. The biggest contribution of the two terms is at zeroth and first order in $k$, i.e. in terms of scalar quantities, we have $E\sim V\pm itk$ to linear order. Therefore the intrinsic error of the three-band approximation is roughly given by 
\begin{equation}
    \frac{V^2+t^2k^2}{6.5\text{eV}}\simeq (1.5+400k^2)\text{meV}.
\end{equation}
This means that if we want a precision on the order of 1 meV, we find that the approximation holds until $k\sim0.1$, so about a tenth of the BZ. This is indeed what we find when we compare the band structure with and without the off-diagonal blocks. More precisely, the confinement energy globally shifts every band by 1 to 2 meV while the imaginary part of the kinetic term breaks the isotropy of the band structure obtained within the low-energy model and gives rise to the $\mathcal{C}_3$ symmetric structure of ellipses seen in experimental studies (see Ref. \cite{rodel_orientational_2014} for example). So the validity of the low-energy model is restrained to the first tenth of the BZ around the $\Gamma$ point. Knowing this, what is the natural order of expansion we can do to the low-energy model ? The relevant terms will be the ones above or around our precision of a few meVs. For the orbital mixing term, the first two corrections are linear and cubic in $k$. We then have $c_0k\sim4$ meV and $c_0k^3\sim0.04$ meV for $k\sim0.1$, so we only take the linear term. As for the kinetic term, the first two corrections are of order $tk^2$ and $tk^4$. This gives $tk^2\sim10$ meV and $tk^4\sim0.1$ meV, we therefore only keep the quadratic term. In conclusion, we can thus expand our three-band model to quadratic order while being coherent with the three-band approximation.

\section{Quadratic expansion of $H_3$}
\label{annexe DL quad H3}
Here, we derive the quadratic expansion of $H_3$ in Eq. (\ref{H3quad}). We remind the reader that the matrices are written in the trigonal basis. For the orbital mixing term, we have
\begin{equation}
    H_{\text{om}}=\begin{pmatrix}
        0&0&ick_x\\
        0&0&ick_y\\
        -ick_x&-ick_y&0
    \end{pmatrix}+\mathcal{O}(k^3),
\end{equation}
with $c=(3/\sqrt{2})c_0$. For the kinetic term, we have
\begin{widetext}
    \begin{equation}
    \text{Re}(H_{\text{cin}})=-t(2+r)\bigg(1-\frac{1}{4}k^2\bigg)\Lambda_0+t\begin{pmatrix}
        -\frac{1}{8}(1-r)(k_x^2-k_y^2)&-\frac{1}{4}(1-r)k_xk_y&\frac{1}{2\sqrt{2}}(1-r)k_xk_y\\
        -\frac{1}{4}(1-r)k_xk_y&\frac{1}{8}(1-r)(k_x^2-k_y^2)&\frac{1}{4\sqrt{2}}(1-r)(k_x^2-k_y^2)\\
        \frac{1}{2\sqrt{2}}(1-r)k_xk_y&\frac{1}{4\sqrt{2}}(1-r)(k_x^2-k_y^2)&0
        \end{pmatrix}+\mathcal{O}(k^4),
\end{equation}
\end{widetext}
where we separated the traceful and traceless parts using 
\begin{equation}
    \begin{pmatrix}
        a&0&0\\
        0&b&0\\
        0&0&c
    \end{pmatrix}=\frac{a+b+c}{3}\Lambda_0+\frac{a-b}{2}\Lambda_3+\frac{a+b-2c}{2\sqrt{3}}\Lambda_8,
\end{equation}
in terms of the Gell-Mann matrices. Now, if we define $t_{\text{eff}}=(t-t_d)/8$, we indeed find Eq. (\ref{H3quad}), neglecting the quadratic terms whenever there already exists a linear term. 
\newpage

\section{Band structure of Eq. (4)} \label{annexe_band_struc_TB}

\begin{widetext}

    \begin{figure}[ht!]
        \includegraphics[width=\textwidth]{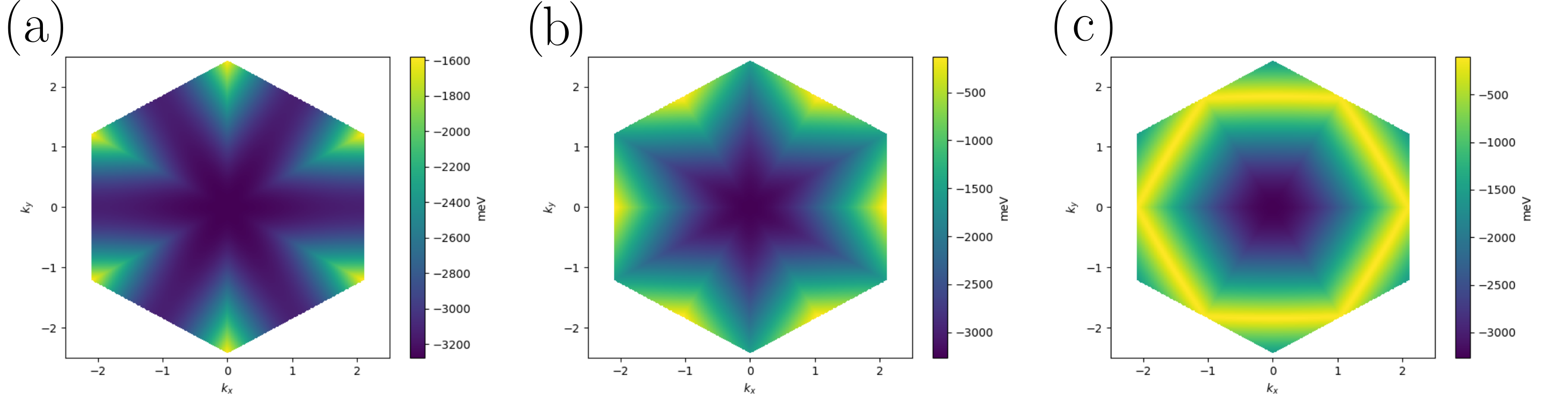}
        \caption{Lower triplet of the TB model in Eq. (\ref{six band matrix}). \textbf{(a)} Lowest band in the full BZ. \textbf{(b)} Second lowest band in the full BZ. \textbf{(c)} Third lowest band in the full BZ.}
        \label{Plots_bandes_TB}
    \end{figure}
    
\end{widetext}

An important issue for our discussion here, from the electronic point of view, is the absence of other electron or hole pockets in the first BZ that would provide other and possibly different metallic and superconducting properties of the system. We therefore plot the dispersion of the lowest three bands in the entire BZ (see Fig. \ref{Plots_bandes_TB}), where we clearly see that the band minima are situated at the $\Gamma$ point and that there are no other pockets in the BZ for our range of chemical potential. Note that Fig.\ref{Plots_bandes_TB} was obtained using the spinless TB Hamiltonian, but the result also holds for the spinful TB Hamiltonian, there is no electron pocket other than the one we consider at the $\Gamma$ point.

\section{Derivation of the spinful two-band low-energy model}
\label{annexe spinful low-energy}
We begin things at Eq.(\ref{intermédiaire spinful low-energy}),
\begin{equation}
    \Tilde{H}_6=\sigma_0\otimes H_3+H_{\text{SOC}},
\end{equation}
where we also apply $U_s=(\sigma_0-i\sigma_y)/\sqrt{2}$ on the spin part, so that $H_{\text{SOC}}$ reads
\begin{widetext}
\begin{equation}
    H_{\text{SOC}}=i\lambda\sigma_x\otimes\begin{pmatrix}
        0&-\frac{1}{\sqrt{3}}&\frac{2}{\sqrt{6}}\\
        \frac{1}{\sqrt{3}}&0&0\\
        -\frac{2}{\sqrt{6}}&0&0
    \end{pmatrix}+i\lambda\sigma_y\otimes\begin{pmatrix}
        0&\frac{1}{\sqrt{3}}&\frac{1}{\sqrt{6}}\\
            -\frac{1}{\sqrt{3}}&0&\frac{1}{\sqrt{2}}\\
            -\frac{1}{\sqrt{6}}&-\frac{1}{\sqrt{2}}&0
    \end{pmatrix}-i\lambda\sigma_z\otimes\begin{pmatrix}
        0&\frac{1}{\sqrt{3}}&\frac{1}{\sqrt{6}}\\
            -\frac{1}{\sqrt{3}}&0&-\frac{1}{\sqrt{2}}\\
            -\frac{1}{\sqrt{6}}&\frac{1}{\sqrt{2}}&0
    \end{pmatrix}.
\end{equation}
\end{widetext}
We then use that fact that close to the $\Gamma$ point, the exact eigenbasis should not be far from the one exactly at the $\Gamma$
point. Consequently, we exactly diagonalize $\Tilde{H}_6(\Gamma)$, for the particular case $\lambda=3d$. This choice is in line with the experimentally determined values of the parameters and happens to greatly simplify the resulting expressions. We thus determine the eigenstates of the two lowest energies, which
we denote as $|u_1\rangle$ and $|u_2\rangle$. We then define a two-band Hamiltonian $H_2$ in the latter's basis, i.e.
\begin{equation}
    H_2=\sum_{i,j=1}^2\langle u_i|\Tilde{H}_6|u_j\rangle|u_i\rangle\langle u_j|
\end{equation}
This way we get to Eqs.(\ref{eq:lowenergy:h0},\ref{eq:lowenergy:h}).

\section{Calculation of $D_{\text{conv}}$ for a general quadratic and isotropic band dispersion.}
\label{annexe calcul Dconv}
 Let us consider a general isotropic and quadratic band $\epsilon=\epsilon_0+\alpha k^2$. Then, we can readily show that $D^{\text{conv}}_{xx}=D^{\text{conv}}_{yy}=D_{\text{conv}}$ and $D^{\text{conv}}_{xy}=0$. We also see that $\mathscr{S}_{\text{occ}}(\mu)=B\big(0;\sqrt{\mu/\alpha}\big)$. We then have
 \begingroup
 \allowdisplaybreaks
 \begin{align}
    D_{\text{conv}}&=\frac{1}{2}\int_{\mathscr{S}_{\text{occ}}(\mu)}\frac{\Delta^2}{\big[\Delta^2+(\alpha k^2-\mu)^2\big]^{3/2}}4\alpha^2k^2 \mathcal D_2 \mathbf{k}\nonumber\\
    &\propto\frac{2\alpha^2}{(2\pi)^2}\int_0^{2\pi}\int_0^{\sqrt{\mu/\alpha}}\frac{\Delta^2}{\big[\Delta^2+(\alpha k^2-\mu)^2\big]^{3/2}}k^3dkd\theta\nonumber\\
    &=\frac{2\alpha^2}{2\pi}\int_0^{\sqrt{\mu/\alpha}}\frac{\Delta^2}{\big[\Delta^2+(\alpha k^2-\mu)^2\big]^{3/2}}k^3dk\nonumber\\
    &=\frac{\alpha^2}{\pi}\frac{1}{2\alpha^2}\int_0^\mu\frac{\Delta^2}{\big[\Delta^2+(\epsilon-\mu)^2\big]^{3/2}}\epsilon d\epsilon\nonumber\\
    &=\frac{1}{2\pi}\Big(\sqrt{\Delta^2+\mu^2}-\Delta\Big).
\end{align}
\endgroup
\end{appendix}
\end{document}